\documentclass[a4paper,11pt]{article}
\pdfoutput=1

\usepackage{jheppub} 
\usepackage[T1]{fontenc} 
\usepackage{amssymb}
\usepackage{amsmath}
\usepackage{color}
\usepackage{xspace}
\usepackage{appendix}
\usepackage{float}
\usepackage{hyperref}
\usepackage{diagbox}
\usepackage[utf8x]{inputenc}
\usepackage{soul}
\usepackage[compat=1.1.0]{tikz-feynhand}
\usepackage{subfig}

\newcommand{\nn}{\nonumber}
\newcommand{\bea}{\begin{align}}
\newcommand{\eea}{\end{align}}
\newcommand{\beq}{\begin{equation}}
\newcommand{\eeq}{\end{equation}}
\newcommand{\bqa}{\begin{eqnarray}}
\newcommand{\eqa}{\end{eqnarray}}

\newcommand{\M}{\ensuremath \text{M}}

\newcommand{\ep}{\epsilon}

\def\ksl{k\!\!\!\slash}

\def\epsl{\epsilon\!\!\!\slash}
\allowdisplaybreaks[4]
\newcommand{\minus}{\ensuremath \scalebox{0.5}[1.0]{\( - \)}}
\newcommand{\pminus}{\hphantom{\minus}}

\title{Analytic  two-loop amplitudes for $tW$ production: leading color and light fermion-loop contributions }
\author[a]{Long-Bin Chen,}
\author[b]{Liang Dong,}
\author[b]{Hai Tao Li,}
\author[c,d,e]{Zhao Li,}
\author[b]{Jian Wang,}
\author[b]{and Yefan Wang}

\affiliation[a]{School of physics and materials science, Guangzhou University, Guangzhou 510006, China}
\affiliation[b]{School of Physics, Shandong University, Jinan, Shandong 250100, China}
\affiliation[c]{Institute of High Energy Physics, Chinese Academy of Sciences, Beijing 100049, China}
\affiliation[d]{School of Physics Sciences, University of Chinese Academy of Sciences, Beijing 100039, China}
\affiliation[e]{Center of High Energy Physics, Peking University, Beijing 100871, China}

\emailAdd{chenlb@gzhu.edu.cn}
\emailAdd{liang.dong@mail.sdu.edu.cn}
\emailAdd{haitao.li@sdu.edu.cn}
\emailAdd{zhaoli@ihep.ac.cn}
\emailAdd{j.wang@sdu.edu.cn}
\emailAdd{wangyefan@sdu.edu.cn}

\abstract{We present the analytical results of the two-loop amplitudes for hadronic $tW$ production, focusing on the leading color and light fermion-loop contributions.
The calculation of the two-loop integrals is performed using the method of canonical differential equations.
The results have been expressed in terms of multiple polylogarithms and checked by comparing the infra-red divergences with the predictions from anomalous dimensions.
Combined with the one-loop squared amplitudes we have computed previously, we obtain the hard function relevant to a NNLO Monte Carlo calculation.
We find that the hard function varies slowly in the region with small top quark velocity but increases dramatically in the region with very large top quark velocity.
After phase space integration, the leading color  hard function gives an about $5.4\%$ correction to the leading order cross section, 
while the light fermion loop contributes about $-1.4\%$.
}

\begin{document}

\maketitle
\flushbottom

\section{Introduction}
\label{sec:intro}

The single top production can be used to study the electroweak interaction of the top quark, which is  important to the precision test of the Standard Model (SM) and the search for new physics beyond the SM. 
In particular, the associated production of a top quark with a $W$ boson is sensitive to $Wtb$ coupling, which has drawn a lot of attention in the community. Recently, precision measurement of the inclusive and differential cross section of this process has been performed by both ATLAS and CMS collaborations at the LHC with $\sqrt{s}=13$ TeV~\cite{ATLAS:2016ofl,ATLAS:2017quy,CMS:2018amb,CMS:2021vqm}. More precise experimental results will be available in the near future after the launch of the Large Hadron Collider (LHC) Run-3.

Besides the precision measurements, high precision theoretical predictions are indispensable in extracting useful information from experimental data. At hadron colliders, the QCD corrections are often significant in making reliable predictions, such as reducing the scale uncertainties and modelling the real process more properly. For $tW$ production, the next-to-leading order (NLO) correction was obtained more than twenty-five years ago~\cite{Giele:1995kr}, and has also been investigated  later by~\cite{Zhu:2001hw,Campbell:2005bb, Cao:2008af, Kant:2014oha}. 
Much effort has been devoted to the studies on the effects beyond NLO QCD corrections, e.g.,  the expansion to next-to-next-to-next-to-leading order in the threshold limit~\cite{Kidonakis:2006bu,Kidonakis:2010ux,Kidonakis:2016sjf,Kidonakis:2021vob} and the all order threshold resummation~\cite{Li:2019dhg}.  To provide the kinematic distributions with higher-order QCD effects, refs.~\cite{Frixione:2008yi,Re:2010bp,Jezo:2016ujg} explored parton showers interfaced to the NLO cross section.

Unfortunately, the next-to-next-to-leading order (NNLO) QCD prediction of $tW$ production  has not been obtained yet. A full NNLO correction consists of the double-real, the real-virtual and the double-virtual contributions. 
In the double-real and the real-virtual corrections, it is necessary to define a scheme to clearly distinguish the process of $tW$ production from that of top-quark pair production.
This issue has been discussed at NLO \cite{Demartin:2016axk},
and deserves a detailed investigation at NNLO.
However, this topic is beyond the scope of this paper,
which aims at only the double virtual contribution.
At the cross section level, the double-virtual part contains the one-loop squared amplitude and the interference between two-loop and tree-level amplitudes. 
The former has been computed analytically in our previous work ~\cite{Chen:2022ntw}. 
In this work we will present the dominant contribution to the double-virtual correction, i.e., the leading-color and the light fermion-loop results.

This paper is organised as follows. 
In section~\ref{sec:bare} we describe the basic setup and the details in the  calculation of two-loop bare amplitudes. 
We discuss the procedure to deal with the ultra-violet (UV) and infra-red (IR) divergences in the bare amplitude in section~\ref{sec:div}. 
The finite part of the squared amplitude is defined as
the hard function which could be used in a NNLO Monte Carlo calculation.  
The numerical results for the leading color and light fermion-loop contributions to the NNLO hard function are presented in section~\ref{sec:num}. 
Finally we conclude in section~\ref{sec:concl}.

\section{Two-loop calculation}
\label{sec:bare}

\subsection{Kinematics and notations}
\label{sec:kinematics}
The $tW$ associated production $g(k_1)+b(k_2)\rightarrow W(k_3)+t(k_4)$ possesses two different massive external particles with $k_3^2 = m_W^2$ and $k_4^2 = (k_1+k_2-k_3)^2=m_t^2$. 
The initial particles are massless, i.e., $k_1^2=k_2^2=0$.
The Mandelstam variables are defined by
\begin{align}
	s=(k_1+k_2)^2\,, \qquad t=(k_1-k_3)^2\,, \qquad u=(k_2-k_3)^2 \,,
\end{align}
which have the relation $s+t+u=m_W^2+m_t^2$. 

The tree-level scattering amplitude is given by
\beq \label{eq:mlo}
  \mathcal{M}^{(0)}  = \frac{e\  g_s \ t_{4,2}^{a} }{\sqrt{2} \sin\theta_W}
  \left(
  \frac{\bar{u}(k_4)\epsl_3^* P_L (\ksl_3+\ksl_4)\epsl_1 u(k_2)}{s}
  +
    \frac{\bar{u}(k_4)\epsl_1 (\ksl_2-\ksl_3+m_t)\epsl_3^* P_L u(k_2) }
  { u-m_t^2} 
  \right),
\eeq 
where $t_{4,2}^{a}$ is the $SU(3)$ generator 
with color indices in the subscript corresponding to the top quark and the bottom quark.
The polarization vectors for the gluon and the $W$ boson are denoted by $\epsilon_1^\mu$ and $\epsilon_3^{*\mu}$, respectively.
$P_L=(1-\gamma_5)/2$ is the left-handed projection operator.

We do not consider the decay of the top quark and the $W$ boson at the moment,
and thus we sum over the polarization states of the gauge bosons in the 
amplitude squared.
Specifically, we apply the equation
\begin{align}
	\sum_i \epsilon_i^{*\mu}(k_3)\epsilon_i^{\nu}(k_3) = -g^{\mu\nu} + \frac{k_3^{\mu} k_3^{\nu}}{m_W^2}.
	\label{eq:pol-sum1}
\end{align} 
for the $W$ boson,
and 
\begin{align}
	\sum_i \epsilon_i^{\mu}(k_1)\epsilon_i^{*\nu}(k_1) = -g^{\mu\nu}+\frac{n^{\mu}k_1^{\nu}+ n^{\nu}k_1^{\mu} }{n\cdot k_1},
	\label{eq-pol-sum2}
\end{align}
for the gluon.
We have chosen a physical gauge for the gluon,
so we do not need to consider the contribution from ghost particles in the external states.
In practice, we can simply neglect the second term in the above equation because of the Ward identity.

\begin{figure}[ht]
    \centering
	\begin{minipage}{0.2\linewidth}
		\centering
		\includegraphics[width=0.7\linewidth]{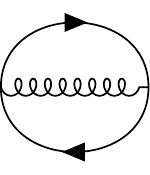}
		\caption*{LO}
	\end{minipage}
		\begin{minipage}{0.2\linewidth}
		\centering
		\includegraphics[width=0.7\linewidth]{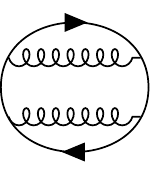}
		\caption*{NLO}
	\end{minipage}
		\begin{minipage}{0.2\linewidth}
		\centering
		\includegraphics[width=0.7\linewidth]{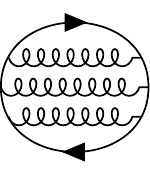}
		\caption*{NNLO}
	\end{minipage}
		\begin{minipage}{0.2\linewidth}
		\centering
		\includegraphics[width=0.7\linewidth]{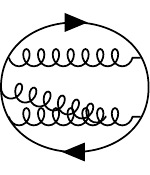}
		\caption*{NNLO}
	\end{minipage}
	\caption{Typical color structure of the squared amplitudes at LO, NLO and NNLO. }
\label{fig:color_examples}
\end{figure}

In this work, we calculate the interference 
between the two-loop, denoted by $\mathcal{M}^{(2)}$,
and tree-level amplitudes,
which can be decomposed  according to the color and flavor structures,
\bqa \label{eq:amp}
\mathcal{A}^{(2)}=\sum_{\text{spins}}|\mathcal{M}^{(0)*} \mathcal{M}^{(2)}|&=&N_c^4 A+ N_c^2 B+ C+\frac{1}{N_c^2}D + n_l\left(N_c^3 E_l+N_c F_l +\frac{1}{N_c}G_l \right)\nonumber\\
&+&n_h\left(N_c^3 E_h+N_c F_h +\frac{1}{N_c}G_h \right) \,,
\label{eq:colorbudget}
\eqa
where $n_l$ and $n_h$ are the total number of light and heavy quark flavors, respectively. $A,B,C,D,E,F,G$ are the corresponding coefficients. 
Notice that we do not perform color average in the squared amplitude.
Therefore, the leading contribution is proportional to $N_c^4$ (or $ N_c^3$ for fermion loop contributions).
The following terms are suppressed by $N_c^2$ in sequence.
The fact that color-summed squared amplitudes can be expanded in a series of $1/N_c^2$, rather than $1/N_c$, has been discussed in detail at one-loop level in \cite{Bern:1990ux}.
In our case, this expansion pattern can be understood.
The $Wtb$ vertex does not affect the color flow and thus can be omitted in the analysis of color structure\footnote{This means that the quark flavor does not change along the fermion line.}.
As a result, the structure of a color-summed squared amplitude is represented by the vacuum graph; see a few examples in figure \ref{fig:color_examples}.
In analysing the color factors of squared amplitudes, the four-gluon vertex in the graph is replaced by two three-gluon vertices,
and any three-gluon vertex is substituted by a difference of two fermion loops due to the identity
\begin{align}
    if^{abc} =2\left( {\rm Tr}[t^a t^b t^c] - {\rm Tr}[t^a t^c t^b]  \right)
    \label{eq:fabc}
\end{align}
Now all the color structure in the graph is a combination of traces of multiple color generators $t^a$.
Applying the $SU(N_c)$ Fierz identity
\begin{align}
    t^a_{ij} t^a_{kl}= \frac{1}{2}\delta_{il}\delta_{kj}-\frac{1}{2N_c}\delta_{ij}\delta_{kl}
    \label{eq:fierz}
\end{align}
to contract the color indices carried by  each gluon propagator,
the resulting color structure consists of only $\delta_{ii}$, each of which contributes a factor of $N_c$
\footnote{This can be illustrated by decomposing every vacuum graph to several fermion loops after replacing each gluon with either a quark-anti-quark  pair or simply dropping it.}.
The leading color (or color planar) contribution is obtained by taking only the first term in the above eq. (\ref{eq:fierz}).
The second term not only contains a factor of $1/N_c$, but also reduces the number of $\delta_{ii}$ 
(or fermion loops in a graph) due to the color flow topology.
When there is a three-gluon vertex,
one  term in eq.(\ref{eq:fabc}) gives rise to the leading color contribution, while the other generates a graph with two fewer fermion loops.
In any case, the expansion is in a series of $1/N_c^2$.
In QCD, this factor is of the same magnitude as the strong coupling,
and serves as a good perturbative expansion parameter.
In addition, the result at each order in this expansion is gauge invariant, and could be calculated independently.
Due to the simple topology of the Feynman diagrams contributing to the leading color \footnote{Here the terminology ``topology'' includes the information on the masses of internal propagators.},
the calculation of this part is notably easier than the full two-loop corrections.
Therefore, we constrain ourselves to present the result at leading color in this paper, i.e., the coefficient $A$ in eq.(\ref{eq:colorbudget}).
As the number of light quarks in the loop is not small compared to $N_c$ and the needed master integrals are almost the same, we also provide the result proportional to $n_l$. 
In summary, we have calculated the following gauge invariant contribution to two-loop squared amplitudes  
\bqa
\mathcal{A}^{(2)}_{{\rm L.C.}+n_l}&\equiv &N_c^4 A+ n_l\left(N_c^3 E_l+N_c F_l +\frac{1}{N_c}G_l \right)\,.
 \label{eq:LC}
\eqa

\subsection{Bare two-loop amplitudes}
\begin{figure}
    \centering
    \begin{minipage}{0.3\linewidth}
		\centering
		\includegraphics[width=0.9\linewidth]{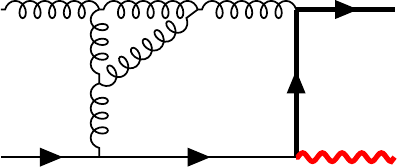}\vspace{15pt}
	\end{minipage}
	    \begin{minipage}{0.3\linewidth}
		\centering
		\includegraphics[width=0.9\linewidth]{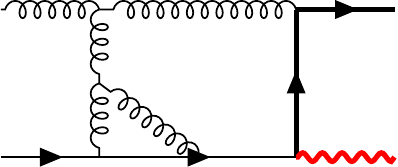}\vspace{15pt}
	\end{minipage}
	    \begin{minipage}{0.3\linewidth}
		\centering
		\includegraphics[width=0.9\linewidth]{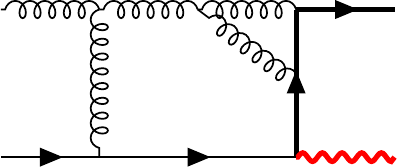}\vspace{15pt}
	\end{minipage}
	    \begin{minipage}{0.3\linewidth}
		\centering
		\includegraphics[width=0.9\linewidth]{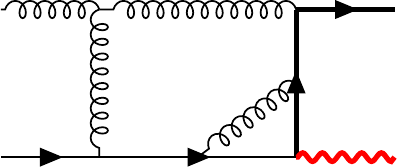}\vspace{15pt}
	\end{minipage}
	    \begin{minipage}{0.3\linewidth}
		\centering
		\includegraphics[width=0.9\linewidth]{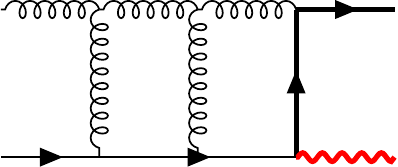}\vspace{15pt}
	\end{minipage}
	    \begin{minipage}{0.3\linewidth}
		\centering
		\includegraphics[width=0.9\linewidth]{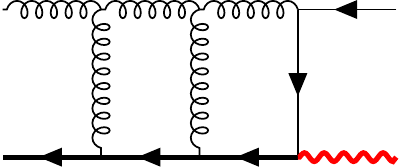}\vspace{15pt}
	\end{minipage}
    \caption{Typical two-loop leading color Feynman diagrams for $tW$ production. The black (red) thick lines stand for the top quark and  the $W$ boson, respectively.}
    \label{lcdiag}
\end{figure}
All the two-loop Feynman diagrams for the process $gb\to Wt$  are generated  
by using {\tt FeynArts} \cite{Hahn:2000kx}.
There are 199 two-loop Feynman diagrams in total, of which 73 diagrams contribute to the leading color and 20 diagrams have a light fermion loop. 
Some typical two-loop Feynman diagrams are displayed in figure \ref{lcdiag}. 
We compute the interference between two-loop and tree-level amplitudes directly, and thus there are no Lorentz indices remaining in the spin-summed result.
The traces of Dirac matrices are performed using the package {\tt FeynCalc} \cite{Shtabovenko:2020gxv}.
We have used conventional dimensional regularization scheme to deal with both the UV and IR divergences, i.e.,
the space-time dimension is extended to $d=4-2\ep$.
The anticommuting $\gamma_5$ scheme  is adopted following ref.\cite{Korner:1991sx}.
The traces containing two $\gamma_5$ matrices are trivial due to $\gamma_5^2=1$ after moving the two $\gamma_5$ matrices adjacent to each other.
The traces with a single $\gamma_5$ matrix is vanishing in our problem because there are only three independent momenta involved.
More detailed discussion can be found in ref.\cite{Chen:2022ntw}.

As a consequence, we obtain the squared amplitude as a linear combination of a large number of scalar Feynman integrals with rational coefficients depending on the kinematic invariants $s,t,u$ and the space-time dimension $d$. 
We find that all the squared amplitudes contributing to the leading color can be expressed in terms of the integrals appearing in the Feynman diagrams shown in figure \ref{lcdiag}.
Explicitly, they are given by
\bqa
I^{\rm L.C.}_{n_1,n_2,\cdots,n_9}=\int\frac{d^d q_1}{i \pi^{d/2}}\frac{d^d q_2}{i \pi^{d/2}}e^{2\gamma_E \epsilon}\frac{D_8^{-n_8} D_9^{-n_9}}{D_1^{n_1} D_2^{n_2} \cdots D_7^{n_7}}.
\eqa
where $q_1$ and $q_2$ are loop momenta and  $D_i$ with $i=1,\cdots,7 $ denote the denominators of the propagators  in each Feynman diagram in figure \ref{lcdiag}
\footnote{Here we abuse the notation of $D_i$. In principle, each Feynman diagram has a set of seven denominators. They are not the same in different diagrams.}.
The other two denominators $D_8$ and $D_9$ are added in order to provide a complete basis for all possible Lorentz invariant scalar products formed by two loop momenta and three external momenta. 

\begin{figure}[ht]
	\centering
	\begin{minipage}{0.32\linewidth}
		\centering
		\includegraphics[width=0.8\linewidth]{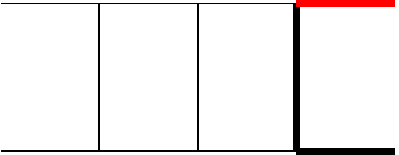}
		\caption*{P1}
	\end{minipage}
	\begin{minipage}{0.32\linewidth}
		\centering
		\includegraphics[width=0.8\linewidth]{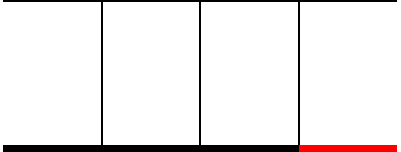}
		\caption*{P2}
	\end{minipage}
\caption{Unfactorized two-loop master integral topologies relevant to the leading color contribution. 
The black (red) thick lines stand for the top quark and the $W$ boson, respectively. 
The other lines denote massless particles.}
\label{fig:P1P2}
\end{figure}

Then all these integrals are reduced to a small set of basis integrals, called master integrals,  using the relations generated by integration by parts (IBP) identities.
We have made use of the package  {\tt FIRE} \cite{Smirnov:2019qkx} in this step.
After considering the symmetry between different topologies, all the two-loop master integrals can be categorized into only two integral families, as displayed in figure \ref{fig:P1P2} \footnote{Here we do not include the  master integrals that can be factorized as two one-loop integrals, which are easy to calculate \cite{Chen:2022ntw}.}.
The denominators for P1 are given by
\begin{align}
D_1&=q_1^2,& D_2&=q_2^2,&D_3&=(q_1-k_1)^2,\nonumber\\
D_4&=(q_1+k_2)^2,& D_5 &=(q_1+q_2-k_1)^2,&
D_6&=(q_2-k_1-k_2)^2,\nonumber\\
D_7&=(q_2-k_3)^2-m_t^2,&
D_8&=(q_1+k_1+k_2-k_3)^2-m_t^2,& D_9&=(q_2-k_1)^2
\end{align}
and  the denominators for P2 are  
\begin{align}
	D_1 &= q_1^2,&
	D_2 &= q_2^2,&
	D_3 &= (q_1-k_2)^2,\nonumber\\
	D_4 &= (q_1-k_3)^2-m_t^2,&
	D_5 &= (q_1 + q_2-k_2)^2,&
	D_6 &= (q_2+k_1)^2,\nonumber\\
	D_7 &= (q_2-k_2+k_3)^2-m_t^2,&
	D_8 &=(q_1-k_1-k_2)^2,&
	D_9 &=(q_2+k_1-k_2)^2.
\end{align}

There are 31 and 38 master integrals in the  P1 and P2 integral families, respectively. 
We have calculated them using the method of differential equations \cite{Kotikov:1990kg,Kotikov:1991pm}.
Taking derivative of one master integral with respective to a kinematic variable, e.g., $s$, the result can be written as a linear combination of the master integrals, since all integrals in a family can be reduced back to the basis.
These differential equations incorporate almost all the information about the integrals.
One can choose a proper kinematic point where the integrals are relatively easier to compute either analytically or numerically, and derive the values at other kinematic points by solving the differential equations.
The latter is made simple if  a canonical basis can be found, i.e.,
the differential equations can be transformed to a form in which the dimensional regulator $\ep$ is decoupled from the kinematic  variables \cite{Henn:2013pwa}.
This is called the $\epsilon$-form or $d\ln$ form. 
A formal solution to this differential equation is given in terms of Chen iterated integrals \cite{Chen:1977oja}.
If the involved symbol letters,
i.e., the arguments of the $d\ln$ form,
are polynomials of the integration variables,
the solution can be  expressed as multiple polylogarithms \cite{Goncharov:1998kja}, which are defined as $G(x)\equiv 1$ and
\bqa
G_{a_1,a_2,\ldots,a_n}(x) &\equiv & \int_0^x \frac{\text{d} t}{t - a_1} G_{a_2,\ldots,a_n}(t)\, ,\\
G_{\overrightarrow{0}_n}(x) & \equiv & \frac{1}{n!}\ln^n x\, .
\eqa
The number of elements in the set $(a_1,a_2,\ldots,a_n)$ is referred to as the transcendental $weight$ of the multiple polylogarithms. 
For the two-loop amplitudes, we need multiple polylogarithms up to transcendental weight four,
and have used {\tt PolyLogTools} \cite{Duhr:2019tlz} to perform efficient numerical evaluation of these functions.
The analytical results of the P1 integral family have been obtained by two of the authors \cite{Chen:2021gjv},
while the master integrals in the P2  family have been calculated in \cite{Long:2021vse},
and also independently checked by
 \cite{Wang:2022xx}.   

\begin{figure}[ht]
	\centering
		\includegraphics[width=0.2\linewidth]{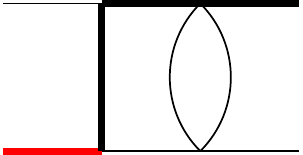}
\caption{An integral topology appearing in the light fermion-loop contribution. The black (red) thick lines stand for the top quark and the $W$ boson, respectively. The other lines denote massless particles.}
\label{fig:topology3}
\end{figure}

\begin{figure}[ht]
	\centering
	\begin{minipage}{0.18\linewidth}
		\centering
		\includegraphics[width=1.0\linewidth]{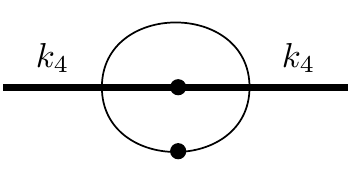}
		\caption*{$\M_{1}$}
	\end{minipage}
	\begin{minipage}{0.18\linewidth}
		\centering
		\includegraphics[width=0.6\linewidth]{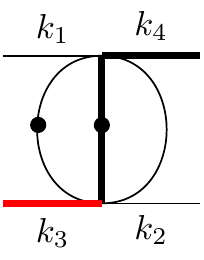}
		\caption*{$\M_{2}$}
	\end{minipage}
	\begin{minipage}{0.18\linewidth}
		\centering
		\includegraphics[width=0.6\linewidth]{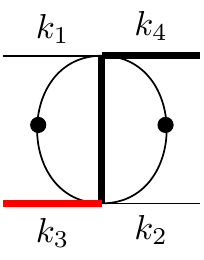}
		\caption*{$\M_{3}$}
	\end{minipage}
	\begin{minipage}{0.16\linewidth}
		\centering
		\includegraphics[width=1.0\linewidth]{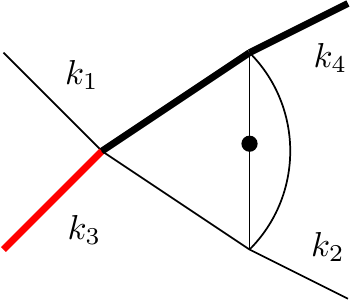}
		\caption*{$\M_{4}$}
	\end{minipage}
		\\
	\begin{minipage}{0.18\linewidth}
		\centering
		\includegraphics[width=1.0\linewidth]{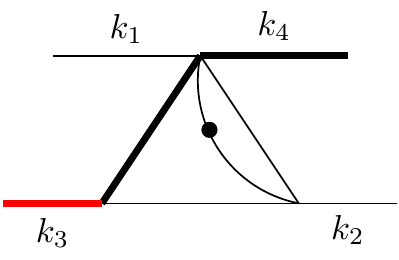}
		\caption*{$\M_{5}$}
	\end{minipage}
	\begin{minipage}{0.18\linewidth}
		\centering
		\includegraphics[width=0.8\linewidth]{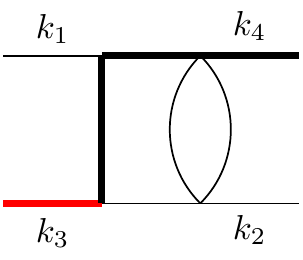}
		\caption*{$\M_{6}$}
	\end{minipage}
	\begin{minipage}{0.18\linewidth}
		\centering
		\includegraphics[width=0.8\linewidth]{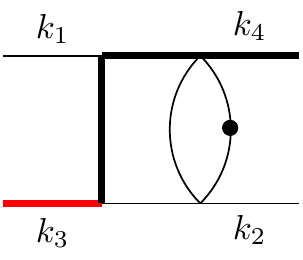}
		\caption*{$\M_{7}$}
	\end{minipage}
\caption{
Topology diagrams for the master integrals defined in eq.(\ref{eq:topology3}).
The black thick (thin) lines stand for the massive (massless) quark and the red thick lines represent the $W$ boson.
The block dot indicates one additional power of the corresponding propagator.}
\label{fig:MIs}
\end{figure}

Then we consider the squared amplitude containing a light fermion loop.
Using the same method as discussed above, we reduce all the scalar integrals to a set of master integrals.
Except those already appearing in the P1 and P2 integral families, another integral family with two massive propagators, as shown in figure \ref{fig:topology3}, should be taken into account.
Explicitly, they are defined as:
\bqa\label{eq:topology3}
& &I_{n_1,n_2,n_3,n_4,n_5,n_6,n_7,n_8,n_9}=\int\frac{d^d q_1}{i \pi^{d/2}}\frac{d^d q_2}{i \pi^{d/2}} e^{2\gamma_E \epsilon}\nonumber\\
& &\frac{[q_2^2]^{-n_6}[(q_2-k_1+k_3)^2]^{-n_7}[(q_2-k_2+k_3)^2]^{-n_8}[(q_1-k_3+k_2)^2]^{-n_9}}{[q_1^2-m_t^2]^{n_1}[(q_1-k_1)^2-m_t^2]^{n_2}[(q_1-k_3)^2]^{n_3}[(q_1+q_2-k_1)^2]^{n_4}[(q_2-k_4)^2]^{n_5}}\,.
\eqa
Since the result of this integral family  is still unknown, we present more details here.
We choose the following canonical basis 
\begin{align}
g_1 &= \epsilon^2 m_t^2 \M_{1}\, ,\nonumber\\
g_2 &= \epsilon^2 u \M_{2}\,,\nonumber\\
g_3 &= \epsilon^2((u-m_t^2)\M_{3}-2m_t^2 \M_{2})\,,\nonumber\\
g_4 &= \epsilon^3(t-m_t^2)\M_{4}\,,\nonumber\\
g_5 &= \epsilon^3(u-m_W^2)\M_{5}\,,\nonumber\\
g_6 &= \epsilon^3(1-2\epsilon)(t-m_W^2)\M_{6}\,,\nonumber\\
g_7 &= \epsilon^3(t-m_t^2)(u-m_t^2)\M_{7}\,,
\end{align}
where
\begin{align}
\M_{1} &=  I_{0,2,0,2,1,0,0,0,0}\, ,\quad&
\M_{2} &=  I_{2,0,0,2,1,0,0,0,0}\,,\quad\nonumber\\
\M_{3} &=  I_{1,0,0,2,2,0,0,0,0}\,,\quad&
\M_{4} &=  I_{0,1,1,1,2,0,0,0,0}\,,\quad\nonumber\\
\M_{5} &=  I_{1,0,1,1,2,0,0,0,0}\,,\quad&
\M_{6} &=  I_{1,1,1,1,1,0,0,0,0}\,,\quad\nonumber\\
\M_{7} &=  I_{1,1,1,1,2,0,0,0,0}\,.
\end{align}
The corresponding topology diagrams are displayed in figure \ref{fig:MIs}.

The differential equations for the canonical basis $\textbf{g}=(g_1,\ldots,g_7)$ can be formulated as 
\bqa
d\, \textbf{g}(x,y,z;\epsilon) = \epsilon(d \Tilde{A})\,  \textbf{g}(x,y,z;\epsilon)
\eqa
with
\bqa
d\, \tilde{A}=\sum_{i=1}^{10} R_i\,  d \ln(l_i)\,.
\eqa
  The letters are given by 
\begin{align}
\begin{alignedat}{2}
l_1 & =x\,,&\quad
l_2 & =x-1\,, \\
l_3 & =y\,, &\quad
l_4 & =y-1\,, \\
l_5 & =z\,,&\quad
l_6 & =z-1\,,\\
l_7 & =x-z \,,&\quad
l_8 & =x\, y -z \, , \\
l_9 & =1+z-x-y\,,&\quad
l_{10} & =y-z\,,
\end{alignedat}
\stepcounter{equation}\tag{\theequation}
\label{alphabet}
\end{align}
with 
\begin{align}
    x=\frac{t}{m_t^2},\quad y=\frac{u}{m_t^2},\quad z=\frac{m_W^2}{m_t^2} \,.
\end{align}
The rational matrices $R_i$ are 
\bqa
R_1 &= \left(
\begin{array}{ccccccc}
 \pminus 0 & \pminus 0 & \pminus 0 & \pminus 0 & \pminus 0 & \pminus 0 & \pminus \pminus 0 \\
\pminus 0 & \pminus 0 & \pminus 0 & \pminus 0 & \pminus 0 & \pminus 0 & \pminus 0 \\
\pminus 0 & \pminus 0 & \pminus 0 & \pminus 0 & \pminus 0 & \pminus 0 & \pminus 0 \\
 \minus 1 & \pminus 0 & \pminus 0 & \pminus 2 & \pminus 0 & \pminus 0 & \pminus 0 \\
\pminus 0 & \pminus 0 & \pminus 0 & \pminus 0 & \pminus 0 & \pminus 0 & \pminus 0 \\
 \minus 1 & \pminus 0 & \pminus 0 & \pminus 2 & \pminus 0 & \pminus 0 & \pminus 0 \\
\pminus 0 & \pminus 0 & \pminus 0 & \pminus 0 & \pminus 0 & \pminus 0 & \pminus 0 \\
\end{array}
\right),~~
R_2&=\left(
\begin{array}{ccccccc}
\pminus 0 & \pminus 0 & \pminus 0 & \pminus 0 & \pminus 0 & \pminus 0 & \pminus 0 \\
\pminus 0 &\pminus 0 & \pminus 0 & \pminus 0 & \pminus 0 & \pminus 0 & \pminus 0 \\
\pminus 0 &\pminus 0 & \pminus 0 & \pminus 0 & \pminus 0 & \pminus 0 & \pminus 0 \\
\pminus 0 &\pminus 0 & \pminus 0 & \minus 4 & \pminus 0 & \pminus 0 & \pminus 0 \\
\pminus 0 &\pminus 0 & \pminus 0 & \pminus 0 & \pminus 0 & \pminus 0 & \pminus 0 \\
\pminus 0 &\pminus 0 & \pminus 0 & \pminus 0 & \pminus 0 & \pminus 0 & \pminus 0 \\
\pminus 0 &\pminus 0 & \pminus 0 & \pminus 0 & \pminus 0 & \pminus 0 & \minus4 \\
\end{array}
\right),\nonumber\\
R_3 &= \left(
\begin{array}{ccccccc}
\pminus 0 & \pminus 0 & \pminus 0 &\pminus 0 & \pminus 0 & \pminus 0 & \pminus 0 \\
\pminus 0 & \pminus 1 & \pminus 0 & \pminus 0 & \pminus 0 & \pminus 0 & \pminus 0 \\
\pminus 0 & \pminus 4 & \pminus 0 & \pminus 0 & \pminus 0 & \pminus 0 & \pminus 0 \\
\pminus 0 & \pminus 0 & \pminus 0 & \pminus 0 & \pminus 0 & \pminus 0 & \pminus 0 \\
\pminus 0 & \minus 1 & \pminus 0 & \pminus 0 & \pminus 0 & \pminus 0 & \pminus 0 \\
\pminus 0 & \pminus 0 & \pminus 0 & \pminus 0 & \pminus 0 & \pminus 0 & \pminus 0 \\
 \pminus 0 & \pminus 0 & \pminus 0 & \pminus 0 & \pminus 0 & \pminus 0 & \pminus 0 \\
\end{array}
\right),~~
R_4 &= \left(
\begin{array}{ccccccc}
\pminus 0 & \pminus 0 & \pminus 0 & \pminus 0 & \pminus 0 & \pminus 0 & \pminus 0 \\
\pminus 0 & \minus 2 & \minus 1 & \pminus 0 & \pminus 0 & \pminus 0 & \pminus 0 \\
\pminus 0 & \minus 4 & \minus 2 & \pminus 0 & \pminus 0 & \pminus 0 & \pminus 0 \\
\pminus 0 & \pminus 0 & \pminus 0 & \pminus 0 & \pminus 0 & \pminus 0 & \pminus 0 \\
\pminus 0 & \pminus 1 & \pminus \frac{1}{2} & \pminus 0 & \pminus 0 & \pminus 0 & \pminus 0 \\
 \minus 2 & \pminus 1 &  \minus \frac{1}{2} & \pminus 0 & \pminus 0 & \pminus 0 & \pminus 0 \\
\pminus 0 & \pminus 0 & \pminus 0 & \pminus 0 & \pminus 0 & \pminus 0 & \minus 4 \\
\end{array}
\right),\nonumber\\
R_5 &= \left(
\begin{array}{ccccccc}
\pminus 0 &\pminus 0 &\pminus 0 &\pminus 0 &\pminus 0 &\pminus 0 &\pminus 0 \\
\pminus 0 &\pminus 0 &\pminus 0 &\pminus 0 &\pminus 0 &\pminus 0 &\pminus 0 \\
\pminus 0 &\pminus 0 &\pminus 0 &\pminus 0 &\pminus 0 &\pminus 0 &\pminus 0 \\
\pminus 0 &\pminus 0 &\pminus 0 &\pminus 0 &\pminus 0 &\pminus 0 &\pminus 0 \\
\pminus 0 &\pminus 1 &\pminus 0 &\pminus 0 &\pminus 2 &\pminus 0 &\pminus 0 \\
\pminus 0 &\pminus 1 &\pminus 0 &\pminus 0 &\pminus 2 &\pminus 0 &\pminus 0 \\
\pminus 0 &\pminus 0 &\pminus 0 &\pminus 0 &\pminus 0 &\pminus 0 &\pminus 0 \\
\end{array}
\right),~~
R_6 &= \left(
\begin{array}{ccccccc}
\pminus 0 &\pminus 0 &\pminus 0 &\pminus 0 &\pminus 0 &\pminus 0 &\pminus 0 \\
\pminus 0 &\pminus 0 &\pminus 0 &\pminus 0 &\pminus 0 &\pminus 0 &\pminus 0 \\
\pminus 0 &\pminus 0 &\pminus 0 &\pminus 0 &\pminus 0 &\pminus 0 &\pminus 0 \\
\pminus 0 &\pminus 0 &\pminus 0 &\pminus 0 &\pminus 0 &\pminus 0 &\pminus 0 \\
\pminus 0 & \minus 1 & \minus \frac{1}{2} &\pminus 0 & \minus 3 &\pminus 0 &\pminus 0 \\
\pminus 0 &\pminus 0 &\pminus 0 &\pminus 0 &\pminus 0 &\pminus 0 &\pminus 0 \\
\pminus 0 & \minus 1 & \minus \frac{1}{2} &\pminus 0 & \minus 3 &\pminus 0 &\pminus 0 \\
\end{array}
\right),\nonumber\\
R_7 &= \left(
\begin{array}{ccccccc}
\pminus 0 &\pminus 0 &\pminus 0 &\pminus 0 &\pminus 0 &\pminus 0 &\pminus 0 \\
\pminus 0 &\pminus 0 &\pminus 0 &\pminus 0 &\pminus 0 &\pminus 0 &\pminus 0 \\
\pminus 0 &\pminus 0 &\pminus 0 &\pminus 0 &\pminus 0 &\pminus 0 &\pminus 0 \\
\pminus 0 &\pminus 0 &\pminus 0 &\pminus 0 &\pminus 0 &\pminus 0 &\pminus 0 \\
\pminus 0 &\pminus 0 &\pminus 0 &\pminus 0 &\pminus 0 &\pminus 0 &\pminus 0 \\
\pminus 0 &\pminus 0 &\pminus 0 &\pminus 0 &\pminus 0 &\pminus 1 &\pminus 0 \\
\pminus 0 &\pminus 1 &\pminus \frac{1}{2} &\pminus 3 &\pminus 3 &\pminus 0 &\pminus 1 \\
\end{array}
\right),~~
R_8&=\left(
\begin{array}{ccccccc}
\pminus 0 &\pminus 0 &\pminus 0 &\pminus 0 &\pminus 0 &\pminus 0 &\pminus 0 \\
\pminus 0 &\pminus 0 &\pminus 0 &\pminus 0 &\pminus 0 &\pminus 0 &\pminus 0 \\
\pminus 0 &\pminus 0 &\pminus 0 &\pminus 0 &\pminus 0 &\pminus 0 &\pminus 0 \\
\pminus 0 &\pminus 0 &\pminus 0 &\pminus 0 &\pminus 0 &\pminus 0 &\pminus 0 \\
\pminus 0 &\pminus 0 &\pminus 0 &\pminus 0 &\pminus 0 &\pminus 0 &\pminus 0 \\
\pminus 1 & \minus\frac{1}{2} &\pminus \frac{1}{4} & \minus\frac{1}{2} & \minus\frac{1}{2} & \minus\frac{1}{2} & \pminus  \frac{1}{2} \\
\pminus 3 & \minus\frac{3}{2} &\pminus \frac{3}{4} & \minus\frac{3}{2} & \minus\frac{3}{2} & \minus\frac{3}{2} & \pminus  \frac{3}{2} \\
\end{array}
\right),\nonumber\\
R_9 &= \left(
\begin{array}{ccccccc}
\pminus 0 &\pminus 0 &\pminus 0 &\pminus 0 &\pminus 0 &\pminus 0 &\pminus 0 \\
\pminus 0 &\pminus 0 &\pminus 0 &\pminus 0 &\pminus 0 &\pminus 0 &\pminus 0 \\
\pminus 0 &\pminus 0 &\pminus 0 &\pminus 0 &\pminus 0 &\pminus 0 &\pminus 0 \\
\pminus 0 &\pminus 0 &\pminus 0 &\pminus 0 &\pminus 0 &\pminus 0 &\pminus 0 \\
\pminus 0 &\pminus 0 &\pminus 0 &\pminus 0 &\pminus 0 &\pminus 0 &\pminus 0 \\
\pminus 1 & \minus\frac{1}{2} &\pminus \frac{1}{4} &\pminus \frac{1}{2} &\pminus \frac{1}{2} & \minus\frac{1}{2} & \minus\frac{1}{2}
   \\
 \minus 3 &\pminus \frac{3}{2} & \minus\frac{3}{4} & \minus \frac{3}{2} & \minus \frac{3}{2} & \pminus \frac{3}{2} &
 \pminus  \frac{3}{2} \\
\end{array}
\right),~~
R_{10}&=\left(
\begin{array}{ccccccc}
\pminus 0 &\pminus 0 &\pminus 0 &\pminus 0 &\pminus 0 &\pminus 0 &\pminus 0 \\
\pminus 0 &\pminus 0 &\pminus 0 &\pminus 0 &\pminus 0 &\pminus 0 &\pminus 0 \\
\pminus 0 &\pminus 0 &\pminus 0 &\pminus 0 &\pminus 0 &\pminus 0 &\pminus 0 \\
\pminus 0 &\pminus 0 &\pminus 0 &\pminus 0 &\pminus 0 &\pminus 0 &\pminus 0 \\
\pminus 0 &\pminus 0 &\pminus 0 &\pminus 0 & \minus 1 &\pminus 0 &\pminus 0 \\
\pminus 0 &\pminus 0 &\pminus 0 &\pminus 0 &\pminus 0 &\pminus 0 &\pminus 0 \\
\pminus 0 &\pminus 0 &\pminus 0 &\pminus 0 &\pminus 3 &\pminus 0 &\pminus 0 \\
\end{array}
\right).
\eqa

The basis integral $g_1$ is simple and can be calculated  directly \cite{Chen:2017xqd},
\beq
 g_1=-\frac{1}{4}-\epsilon^2\frac{5\pi^2}{24}-\epsilon^3\frac{11\zeta(3)}{6}-\epsilon^4\frac{101\pi^4}{480}+{\cal O}(\epsilon^{5}).
\eeq
The boundary condition for $g_3$ is chosen at  $u=0$,
\begin{align}
g_{3}|_{u=0}=1+\epsilon^2\frac{\pi^2}{2}-\epsilon^3\frac{8\zeta(3)}{3}+\epsilon^4\frac{7\pi^4}{40}+{\cal O}(\epsilon^{5}).
\end{align}
The boundary conditions of other basis integrals $g_i$ can be found using the regularity conditions.
We know that the integrals $g_i$ do not contain any branch cut starting at 
the points corresponding to $m_W^2 = 0$, $s = m_t^2$, $ u = 0$, or $ t=0$.
Therefore, the derivatives of the integrals do not have  poles at these points, which can generate relations among the integrals that appear as coefficients of the poles.

With these boundary conditions, it is ready to obtain the analytical results for the canonical basis.
The integration path from the boundary point to the physical point does not cross over any branch cut,
and thus there is no need to perform analytic continuation.
Consequently, the integrals in the above family are real in the relevant physical region.

\section{UV and IR divergences}
\label{sec:div}

\subsection{Renormalization}
After the calculation of two-loop Feynman diagrams, the bare amplitude, denoted by $\mathcal{M}_{\rm bare}$, contains UV and IR divergences. To cancel the UV divergence, we generate the Feynman diagrams with the counter-terms, which arise from the renormalization of the couplings, masses and field strength.  
Then the renormalized QCD amplitude is obtained by 
\begin{align}\label{eq:mren}
    \mathcal{M}_{\rm ren} = Z_{g}^{1/2}Z_{b}^{1/2}Z_{t}^{1/2}\left(\mathcal{M}_{\rm bare}\big|_{\alpha_s^{\rm bare} \to Z_{\alpha_s} \alpha_s;\  m_{t,\rm bare} \to Z_{m} m_t }\right)\;,
\end{align}
where $Z_{g,b,t}$ are the wave function renormalization factors for the external colored particles. The strong coupling $\alpha_s$ and top quark mass are renormalized by the factor $Z_{\alpha_s}$ and $Z_{m}$, respectively.   
In our notation, the amplitude and renormalization factors are expanded as a series of $\alpha_s/4\pi$, e.g.,
$$Z_x = 1 + \frac{\alpha_s}{4\pi} \delta Z_x^{(1)}+ \left(\frac{\alpha_s}{4\pi}\right)^2 \delta Z_x^{(2)}.$$
The renormalized amplitude  is 
\begin{align}
  \mathcal{M}_{\rm ren} &= \mathcal{M}^{(0)}_{\rm bare} + \frac{\alpha_s} {4\pi} ( \mathcal{M}^{(1)}_{\rm bare}+ \mathcal{M}^{(1)}_{\rm C.T.})+\left(\frac{\alpha_s} {4\pi} \right)^2( \mathcal{M}^{(2)}_{\rm bare}+ \mathcal{M}^{(2)}_{\rm C.T.})
  \nn \\
  &= \mathcal{M}_{\rm ren}^{(0)} +\frac{\alpha_s} {4\pi} \mathcal{M}_{\rm ren}^{(1)} +\left(\frac{\alpha_s} {4\pi} \right)^2 \mathcal{M}_{\rm ren}^{(2)}\,
  \label{eq:ampren2}
\end{align}
with the counter-term contribution 
\begin{align} \label{eq:cts}
    \mathcal{M}^{(1)}_{\rm C.T.} =& \delta Z_1 \mathcal{M}^{(0)}_{\rm ren} + \delta Z_m^{(1)}\mathcal{M}^{(0),  m_t}_{\rm C.T.}\;,
    \nn \\ 
    \mathcal{M}^{(2)}_{\rm C.T.} =& \delta Z_2 \mathcal{M}^{(0)}_{\rm ren} + \delta Z_3 \mathcal{M}^{(1)}_{\rm bare} + \delta Z_4 \mathcal{M}^{(0),  m_t}_{\rm C.T.}
    +\left(\delta Z_m^{(1)}\right)^2\mathcal{M}^{\prime(0), m_t}_{\rm C.T.}+ \delta Z_m^{(1)}\mathcal{M}^{(1),  m_t}_{\rm C.T.}\;.
\end{align}
$\mathcal{M}^{(0),  m_t}_{\rm C.T.}$ ($\mathcal{M}^{(1),  m_t}_{\rm C.T.}$) denotes the amplitude with an insertion of a mass counter-term vertex in the tree-level (one-loop)  Feynman diagrams, while $\mathcal{M}^{\prime(0),  m_t}_{\rm C.T.}$ contains two such  mass counter-term vertices in  the tree-level  Feynman diagrams. 
The definitions of $ \delta Z_i$ in eq. (\ref{eq:cts}) are given by
\begin{align}
    \delta Z_1 =&\frac{1}{2} \left[ \delta Z_g^{(1)}+\delta Z_b^{(1)}+\delta Z_t^{(1)}+\delta Z_{\alpha_s}^{(1)}\right],
    \nn \\ 
    \delta Z_2 =&
       \frac{1}{2} \left[ \delta Z_g^{(2)}+\delta Z_b^{(2)}+\delta Z_t^{(2)}+\delta Z_{\alpha_s}^{(2)}\right]
       +\frac{1}{4}\left[   \delta Z_g^{(1)} \delta Z_b^{(1)}+\delta Z_g^{(1)} \delta Z_t^{(1)}
               \nn \right. \\  & \left.
       +\delta Z_b^{(1)} \delta Z_t^{(1)}
       +\delta Z_g^{(1)} \delta Z_{\alpha_s}^{(1)}
       +\delta Z_b^{(1)} \delta Z_{\alpha_s}^{(1)}+\delta Z_t^{(1)} \delta Z_{\alpha_s}^{(1)} \right]
       \nn \\ &
       -\frac{1}{8}\left[ \left(\delta Z_g^{(1)}\right)^2+\left(\delta Z_b^{(1)}\right)^2+\left(\delta Z_t^{(1)}\right)^2+\left(\delta Z_{\alpha_s}^{(1)}\right)^2\right],
    \nn \\ 
    \delta Z_3 =&\delta Z_1 +\delta Z_{\alpha_s}^{(1)} ,
    \nn \\ 
    \delta Z_4 =& \delta Z_1  \delta Z_m^{(1)}    +  \delta Z_m^{(2)}. 
\end{align}
Notice that those quantities in the two lines of  eq. (\ref{eq:ampren2}) should be calculated in $d$-dimensional space-time and kept up to $\mathcal{O}(\ep^0)$. 

We have adopted the on-shell  scheme for the renormalization of the  wave functions and the top-quark mass. 
The strong coupling $\alpha_s$ is renormalized  in the $\overline{\rm MS}$ scheme. Up to two loops, the relevant renormalization constants are given by~\cite{Broadhurst:1991fy,Melnikov:2000zc,Czakon:2007wk,Czakon:2007ej}
\begin{align}
    Z_g=&1
    +\left(\frac{\alpha_s}{4\pi}\right)T_F n_h D_\ep\left(-\frac{4}{3\ep}\right) +\left(\frac{\alpha_s}{4\pi}\right)^2 T_F n_h  D_\ep^2\Biggl[ C_F\left(-\frac{2}{\ep}-15\right)+C_A\left(\frac{35}{9\ep^2}-\frac{5}{2\ep}\right.\nn \\
    &\left.+\frac{13}{12}\right)-\frac{16}{9\ep^2}T_F n_l-\frac{\pi^2}{9}\beta_0+\frac{4}{3}\beta_0\ln\left(\frac{\mu^2}{m_t^2}\right) \left(-\frac{1}{\ep}+\frac{1}{2}\ln\left(\frac{\mu^2}{m_t^2}\right)\right) \Biggr],\nn \\
    Z_b =& 1+\left(\frac{\alpha_s}{4\pi}\right)^2 C_F T_F n_h D_\ep^2 \left(\frac{1}{\ep}  - \frac{5}{6}\right) , \nn \\    
    Z_t=&1+\left(\frac{\alpha_s}{4\pi}\right) C_F D_\ep \left(-\frac{3}{\ep}-4-8\ep-16\ep^2\right) + \left(\frac{\alpha_s}{4\pi}\right)^{2} C_F D_\ep^2\Bigg[T_F n_h \left(\frac{1}{\ep}+\frac{947}{18}-5\pi^2\right) \nn \\
    &+ T_F n_l \left(-\frac{2}{\ep^2}+\frac{11}{3\ep}+\frac{113}{6}+\frac{5\pi^2}{3}\right)+ C_F \bigg(\frac{9}{2\ep^2}+\frac{51}{4\ep}+\frac{433}{8}-13\pi^2 \nn \\
    &+16\pi^2\ln2-24\zeta_3\bigg)+C_A\left(\frac{11}{2\ep^2}-\frac{127}{12\ep}-\frac{1705}{24}+\frac{49\pi^2}{12}-8\pi^2\ln2+12\zeta_3\right) \nn \\
    &+ \beta_0\ln\left(\frac{\mu^2}{m_t^2}\right)\left(-\frac{3}{\ep}-4+\frac{3}{2}\ln\left(\frac{\mu^2}{m_t^2}\right)\right)\Bigg],
    \nn \\
    Z_m=&1
    +\left(\frac{\alpha_s}{4\pi}\right) C_F D_\ep \left(-\frac{3}{\ep}-4-8\ep-16\ep^2\right) 
    + \left(\frac{\alpha_s}{4\pi}\right)^{2} C_F D_\ep^2\Bigg[\beta_0 \left(\frac{3}{2\ep^2}-\frac{5}{4\ep}-\frac{143}{8}\right.
    \nn \\
    &\left.+\frac{7\pi^2}{4}\right)+4 T_F n_l \left(-3+\pi^2\right)+C_F \biggl(\frac{9}{2\ep^2}+\frac{45}{4\ep}+\frac{199}{8}
    -5\pi^2+8\pi^2\ln2-12\zeta_3\biggr)
    \nn \\
    &+C_A\left(-\frac{7}{2\ep}+\frac{77}{4}-6\pi^2-4\pi^2\ln2+6\zeta_3\right) 
    +\beta_0\ln\left(\frac{\mu^2}{m_t^2}\right)\left(-\frac{3}{\ep}-4+\frac{3}{2}\ln\left(\frac{\mu^2}{m_t^2}\right)\right)\Bigg] ,
    \nn \\
    Z_{\alpha_s}=& 1-\frac{\alpha_s}{4\pi}
   \frac{\beta_0}{ \ep}
    +\left(\frac{\alpha_s}{4\pi} \right)^2 \left(\frac{\beta_0^2}{\ep^2}-\frac{\beta_1}{2\ep}\right),
\end{align}
where
\begin{align}
     D_{\ep} & \equiv e^{\gamma_E\ep} \Gamma(1+\ep) \left(\frac{ \mu^2}{m_t^2}\right)^{\ep},
         \nn \\
     \beta_0 & = \frac{11}{3}C_A-\frac{4}{3}T_F (n_l+n_h)\,, 
         \nn \\
    \beta_1 & = \frac{34}{3}C_A^2 - \frac{20}{3} C_A T_F (n_l+n_h)- 4 C_F T_F(n_l+n_h) \,,
\end{align}
with $n_l=5$ and $n_h$=1 for the process $gb\to Wt$.

\subsection{IR divergences}
The renormalized amplitudes in eq. (\ref{eq:ampren2}) still contain IR divergences,
which have a general structure that depends only on the properties of external particles.
These divergences can be factorized from the finite part of the amplitudes because of the properties of the amplitudes in the soft and collinear limits \cite{Catani:1998bh,Becher:2009cu,Gardi:2009qi,Becher:2009qa,Becher:2009kw,Ferroglia:2009ep,Mitov:2010xw}, i.e.,
\begin{align}
  \mathcal{M}_{\rm ren}=  \mathbf{Z} \mathcal{M}_{\rm fin} \;.
\end{align}
where the factor $\mathbf{Z}$  encodes all the IR divergences of the scattering amplitudes 
and has been computed to two loops for a general processes with massive particles.
For single top productions, the IR divergences have been studied at three-loop level  \cite{Kidonakis:2019nqa};
see \cite{Liu:2022elt} for more general processes.

The IR divergences of the amplitudes in QCD can be reproduced by the corresponding amplitudes in the soft-collinear effective theory (SCET).
As such, the IR divergences are transformed into UV ones, since all the loop integrals in the effective theory are scaleless and thus vanish in dimensional regularization.
These UV divergences are closely related to the anomalous dimensions of effective operators corresponding to the relevant process.
For $gb\to Wt$, the anomalous dimension $\mathbf{\Gamma}_h$ up to two loops is given by 
\begin{align} \label{eq:gammah}
    \mathbf{\Gamma}_h =&  \mathbf{T}_1 \cdot \mathbf{T}_2 \gamma_{\rm cusp}(\hat{\alpha}_s)  \ln\frac{\mu^2}{-s}
    +\mathbf{T}_1 \cdot \mathbf{T}_4 \gamma_{\rm cusp}(\hat{\alpha}_s)  \ln\frac{m_t\mu}{m_t^2-u}
         \nonumber \\
    & +\mathbf{T}_2 \cdot \mathbf{T}_4 \gamma_{\rm cusp}(\hat{\alpha}_s)  \ln\frac{m_t\mu}{m_t^2-t}+ \gamma_g(\hat{\alpha}_s)  + \gamma_q(\hat{\alpha}_s) + \gamma_t(\hat{\alpha}_s) 
        \nonumber \\
    = &  \frac{\gamma_{\rm cusp}(\hat{\alpha}_s)}{2} \left(-C_A \ln\frac{\mu^2}{-s} -C_A \ln \frac{m_t\mu }{m_t^2-u} +(C_A-2C_F)\ln\frac{m_t\mu }{m_t^2-t} \right)     \nonumber \\ 
    & + \gamma_g(\hat{\alpha}_s)  + \gamma_q(\hat{\alpha}_s) + \gamma_t(\hat{\alpha}_s)\;,
\end{align}
where $\mathbf{T}_i$ is the color charge associated with the external particle $i$, as  defined in \cite{Catani:1996jh}.
The anomalous dimensions $\gamma_{\rm cusp}$, $\gamma_g$, $\gamma_b$ and $\gamma_t$ are universal quantities in the sense that they are independent of the hard scattering process.
Their two-loop expressions can be found in the appendix of \cite{Li:2013mia}.
Here we have used the notation $\hat{\alpha}_s$ for the strong coupling
with five light quark flavours in its anomalous dimension. 
The matching to the coupling $\alpha_s$ involving heavy quarks in its renormalization is given by  
$\alpha_s = \xi \hat{\alpha}_s$ with \cite{Steinhauser:2002rq,Becher:2009kw}
\begin{align}
    \xi = 1 +  \frac{\alpha_s}{3 \pi}  T_F n_h \frac{D_\epsilon-1}{\epsilon}+\mathcal{O}(\alpha_s^2) \;.
\end{align}

Then the $\mathbf{Z}$ factor can be obtained by  
\begin{align}\label{eq:lnz}
    \ln \mathbf{Z}=& 
    \frac{\alpha_s}{4\pi} \frac{1}{\xi}\left[ \frac{\mathbf{\Gamma}_h^{\prime(0)}}{4\epsilon^2} +\frac{\mathbf{\Gamma}_h^{(0)}}{2 \epsilon}  \right]
         \nonumber \\
    & + \left(\frac{\alpha_s}{4\pi}\right)^2\frac{1}{\xi^2}
    \left[ \frac{-3 \hat{\beta}_0 \mathbf{\Gamma}_h^{\prime(0)}}{16 \epsilon^3} + \frac{\mathbf{\Gamma}_h^{\prime(1)}-4 \hat{\beta}_0 \mathbf{\Gamma}_h^{(0)}}{16 \epsilon^2} + \frac{\mathbf{\Gamma}_h^{(1)}}{4\epsilon}\right]
    + \mathcal{O}(\alpha_s^3)\;,
\end{align}
where $\hat{\beta}_0$ is the LO $\beta$-function for the coupling $\hat{\alpha}_s$ in SCET and thus $\hat{\beta}_0 = \beta_0|_{n_h\to0}$.   We have defined $\mathbf{\Gamma}_h^{\prime}=\partial \mathbf{\Gamma}_h/\partial \ln\mu$\;. 
With all the ingredients at hand it is straightforward to compute $\mathbf{Z}$ in the expansion form
\begin{align}
    \mathbf{Z} = 1 + \frac{\alpha_s}{4\pi}\mathbf{Z}^{(1)} + \left( \frac{\alpha_s}{4\pi} \right)^2 \mathbf{Z}^{(2)}~ + \mathcal{O}(\alpha_s^3).
\end{align}

The finite amplitude up to two loops can be written as
\begin{align}
    \mathcal{M}_{\rm fin} 
     & = \mathcal{M}_{\rm fin}^{(0)} + \frac{\alpha_s}{4\pi} \mathcal{M}_{\rm fin}^{(1)}+\left(\frac{\alpha_s}{4\pi} \right)^2\mathcal{M}_{\rm fin}^{(2)}   \,,
\end{align}
where
\begin{align}
   \mathcal{M}_{\rm fin}^{(0)} &= \mathcal{M}_{\rm ren}^{(0)}\,,
   \nn \\
   \mathcal{M}_{\rm fin}^{(1)} &= \mathcal{M}_{\rm ren}^{(1)}-\mathbf{Z}^{(1)} \mathcal{M}_{\rm ren}^{(0)}\,,
   \nn \\
   \mathcal{M}_{\rm fin}^{(2)} & =  \mathcal{M}_{\rm ren}^{(2)}+ ((\mathbf{Z}^{(1)})^2-\mathbf{Z}^{(2)})\mathcal{M}_{\rm ren}^{(0)}  -\mathbf{Z}^{(1)} \mathcal{M}_{\rm ren}^{(1)} \,.
\end{align}
All the IR divergences are cancelled  order by order on the right hand side of the above equations, which serves as a strong check of our calculations. 

The hard function for $tW$ production in SCET is defined as $\big|  \mathcal{M}_{\rm fin}  \big|^2$. The perturbative expansion of the hard function is
\begin{align}
    H = H^{(0)} + \frac{\alpha_s}{4\pi} H^{(1)}  +\left(\frac{\alpha_s}{4\pi} \right)^2 H^{(2)}\;,
\end{align}
where
\begin{align} \label{eq:hard}
    H^{(0)} = & \Big|\mathcal{M}_{\rm fin}^{(0)} \Big|^2\;,
    \nn \\ 
    H^{(1)} = &  \mathcal{M}_{\rm fin}^{(1)}  \mathcal{M}_{\rm fin}^{(0)*} + \mathcal{M}_{\rm fin}^{(0)}  \mathcal{M}_{\rm fin}^{(1)*} \;,
    \nn \\ 
    H^{(2)} = &  \mathcal{M}_{\rm fin}^{(2)}  \mathcal{M}_{\rm fin}^{(0)*} + \mathcal{M}_{\rm fin}^{(0)}  \mathcal{M}_{\rm fin}^{(2)*} + \Big|\mathcal{M}_{\rm fin}^{(1)} \Big|^2\;.
\end{align}
We have obtained the result of $\Big|\mathcal{M}_{\rm fin}^{(1)} \Big|^2 $ with full color information in~\cite{Chen:2022ntw}.
Similar to eq.~(\ref{eq:amp}), the NNLO corrections to the hard function can be written as
\begin{align}\label{eq:H2}
    H^{(2)} =& N_c^4 H_A + N_c^2 H_B + H_C + \frac{1}{N_c^2} H_D + n_l \left(N_c^3 H_{El}+N_c H_{Fl} + \frac{1}{N_c} H_{Gl} \right)
         \nn \\   + &   n_h \left(N_c^3 H_{Eh}+N_c H_{Fh} + \frac{1}{N_c} H_{Gh} \right)\;.
\end{align}
In this work, we will provide the analytical result of the leading color contribution in eq.~(\ref{eq:H2}), which is defined by
\begin{align}\label{eq:h2lc}
    H^{(2)}_{{\rm L.C.}} \equiv  N_c^4 H_A  \;,
\end{align}
as well as the one including light fermion-loop contribution
\begin{align}\label{eq:h2lcnl}
    H^{(2)}_{{\rm L.C.}+n_l} \equiv  N_c^4 H_A  + n_l \left(N_c^3 H_{El}+N_c H_{Fl} + \frac{1}{N_c} H_{Gl} \right)\;.
\end{align}

\section{Numerical results}\label{sec:num}

When we present the numerical results in this section,  a factor $e^2 g_s^2/\sin^2{\theta_W}$ has been extracted out in the hard functions. 
The ratio between the mass of the $W$ boson and that of the top quark is fixed to be $m_W^2/m_t^2=3/14$ as in ref.~\cite{Chen:2022ntw}.  The renormalization scale $\mu$ is set to be $m_t$. The phase space is parameterized with the velocity $\beta_t$ and polar angle $\theta$ of the final-state top quark  in the center-of-mass frame of the initial partons assuming the incoming gluon moving in the $z$ direction.  The Mandelstam variables can be written as 
\begin{align}
    s=m_W^2-m_t^2+2\Delta\,, \quad
    t=m_t^2-\Delta (1+\beta_t\cos\theta)\,,  \quad
    u=m_t^2-\Delta (1-\beta_t\cos\theta)\,
\end{align}
with $\Delta =  \left(m_t^2+m_t\sqrt{m_W^2+\beta_t^2(m_t^2-m_W^2)}\right) /(1-\beta_t^2)$.  The phase space constraint is $0\le \beta_t<1$ and $-1\le \cos\theta \le 1$~.

\begin{figure}
    \centering
    \includegraphics[width = 0.49 \textwidth]{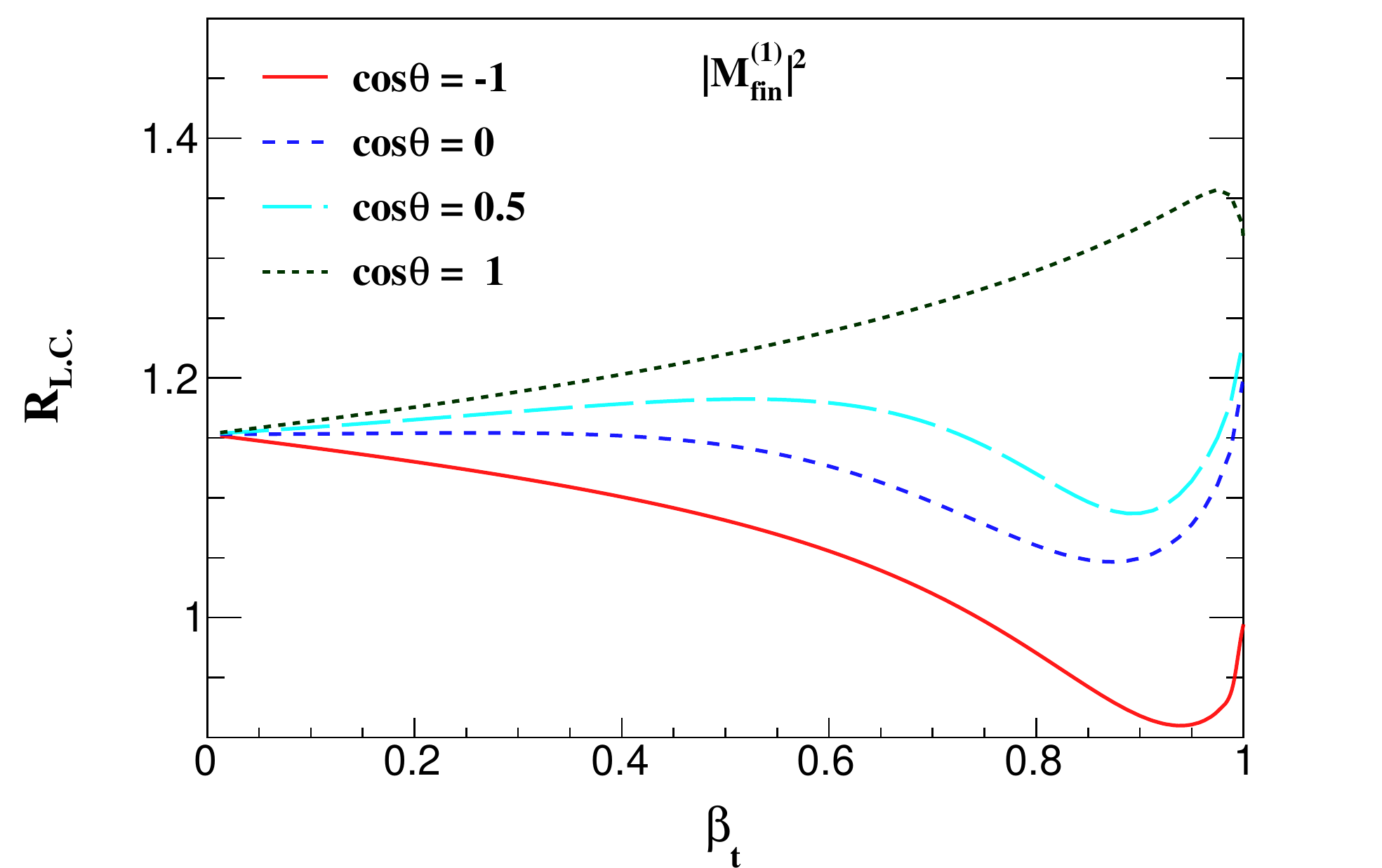}
    \includegraphics[width = 0.49 \textwidth]{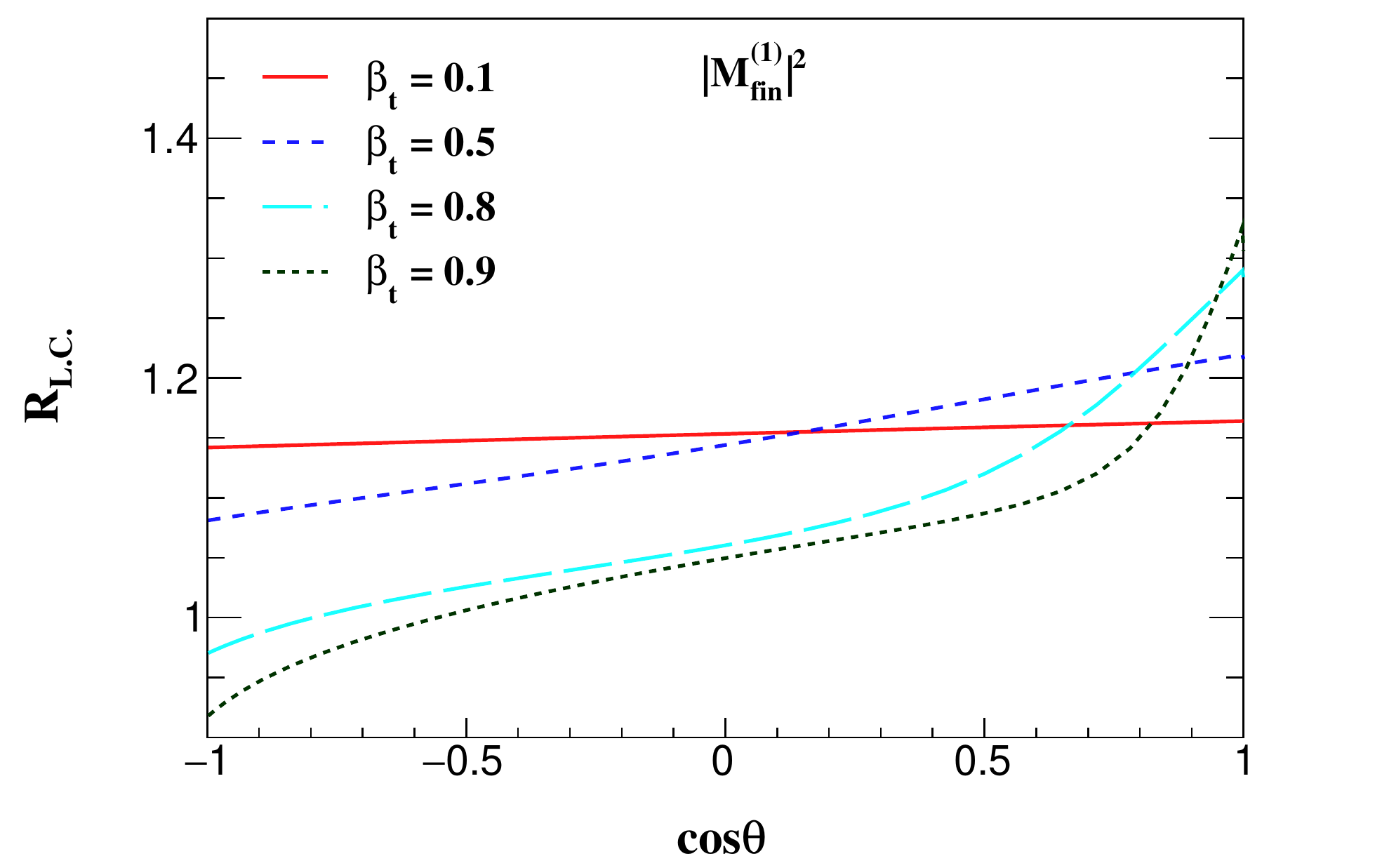}
    \caption{The ratio of the leading color contribution of $\Big|\mathcal{M}_{\rm fin}^{(1)} \Big|^2$ to the one with full color dependence.  The left plot shows the $\beta_t$ distribution with $\cos\theta$ fixed at $-1$ (red), $0$ (blue), $0.5$ (cyan), and $1$ (dark green). The right plot presents the ratio over $\cos\theta$ with $\beta_t$ fixed at $0.1$ (red), $0.5$ (blue), $0.8$ (cyan), and $0.9$ (dark-green). }
    \label{fig:sqLC}
\end{figure}

We present the leading color result of the hard function in this paper
and expect that this represents the most significant contribution, as we argued in section 
\ref{sec:kinematics}.
To illustrate this in practice, we 
examine the  one-loop squared matrix element $\Big|\mathcal{M}_{\rm fin}^{(1)} \Big|^2$,
which is known with full color dependence~\cite{Chen:2022ntw}.
It is convenient to define the ratio 
$R_{\rm L.C.}= \Big|\mathcal{M}_{\rm fin}^{(1), {\rm L.C.}} \Big|^2/\Big|\mathcal{M}_{\rm fin}^{(1)} \Big|^2$, which 
is a function of $\beta_t$ and $\cos\theta$.
Then  $1-R_{\rm L.C.}$ estimates the effect of $1/N_c^2$ suppressed, dubbed sub-leading color, contributions.
The numerical result of $R_{\rm L.C.}$ is presented in figure \ref{fig:sqLC}.
In the small $\beta_t$ region ($\beta_t\le 0.2$), $R_{\rm L.C.}$ is around 1.15, insensitive to the variation of $\cos\theta$.
As the increasing of $\beta_t$, 
the dependence of $R_{\rm L.C.}$ on $\cos\theta$ becomes stronger,
and $R_{\rm L.C.}$ can grow to  1.36 
or decrease to $0.91$ at most.
However,  $|1-R_{\rm L.C.}|$ is less than $20\%$ except for the region with $\beta_t\ge 0.8$ and $\cos\theta\ge 0.8$,
which indicates that the leading color result is the dominant contribution, as expected.

\begin{figure}
    \centering
    \includegraphics[width = 0.49 \textwidth]{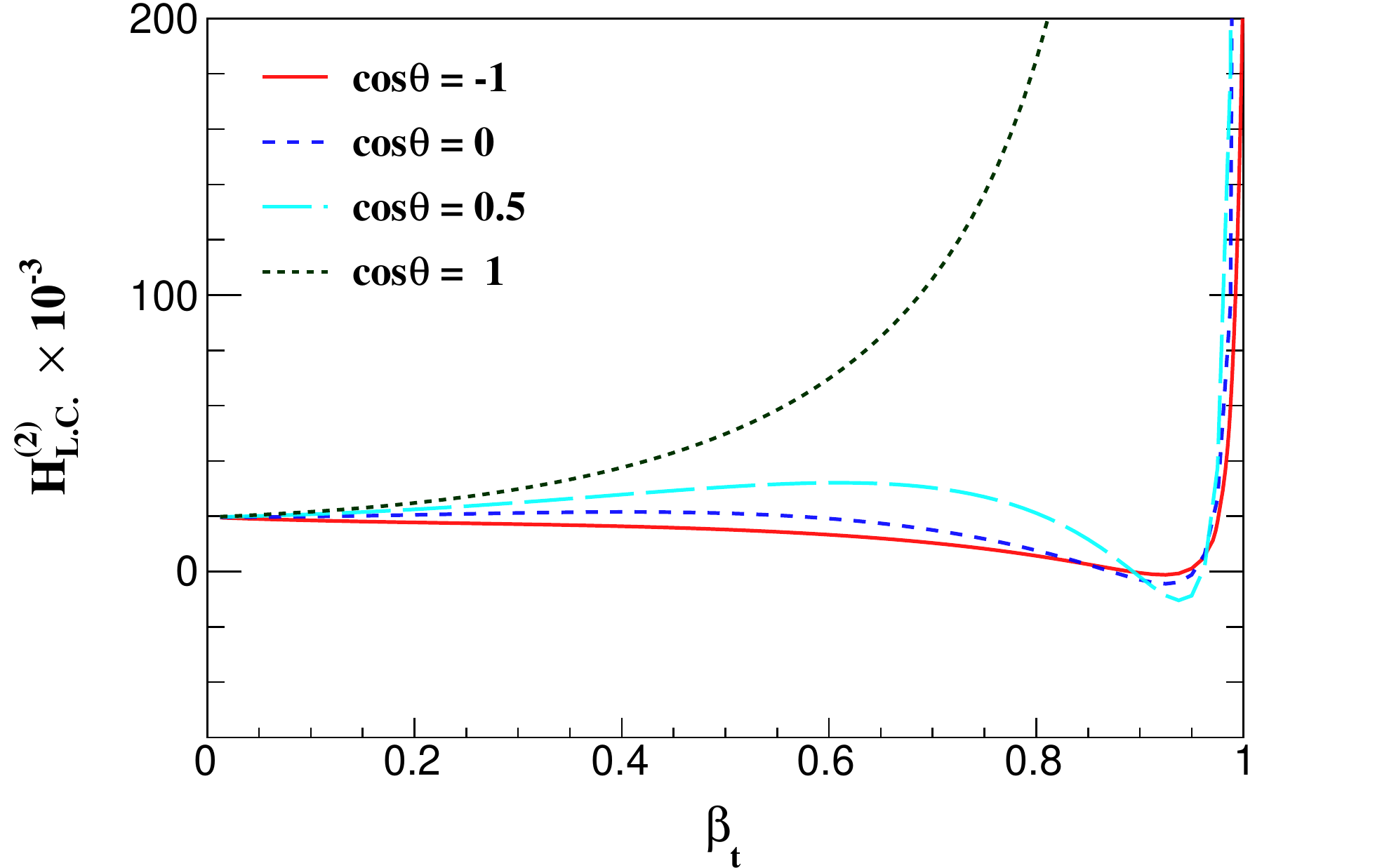}
    \includegraphics[width = 0.49 \textwidth]{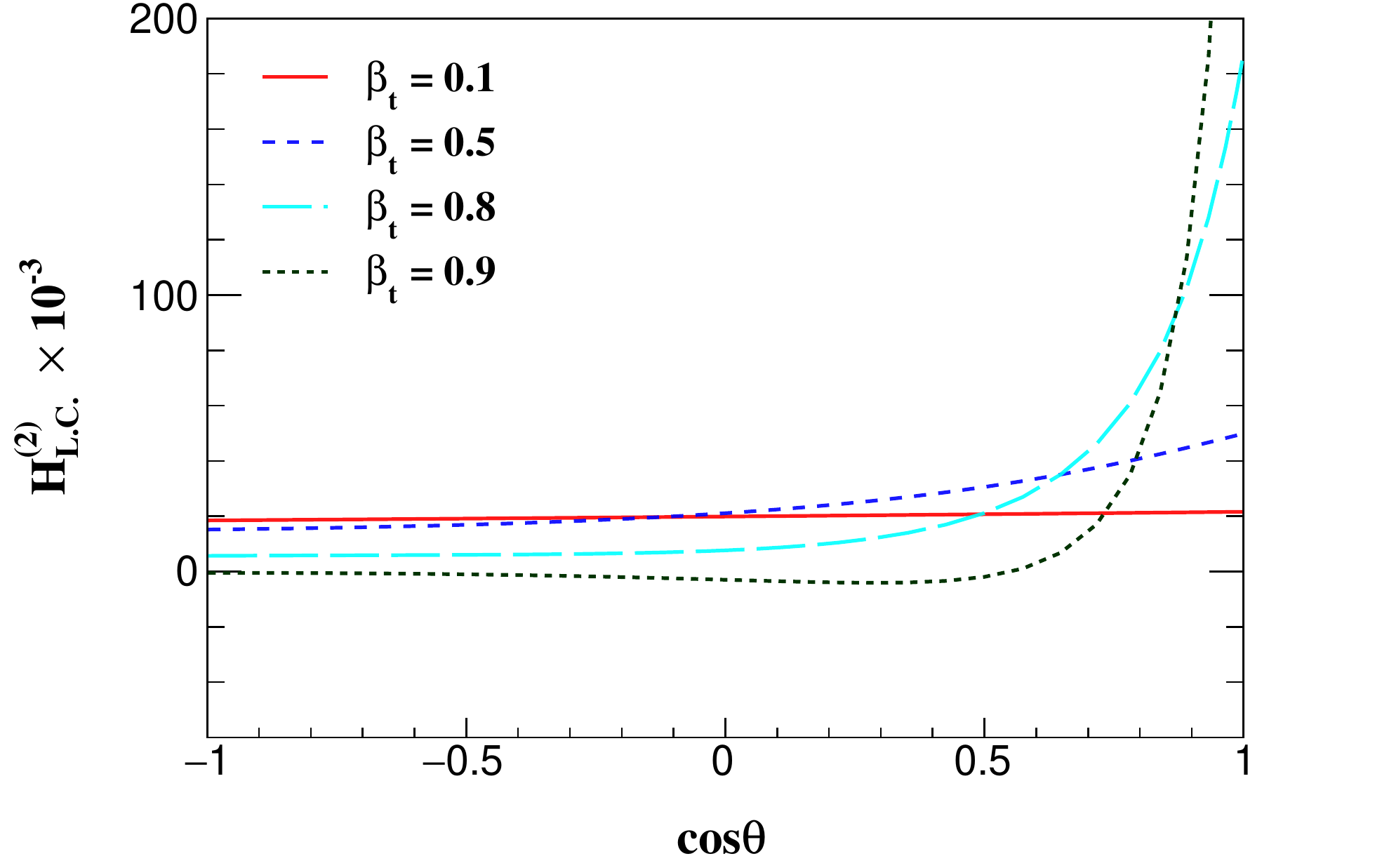}    
    \includegraphics[width = 0.49 \textwidth]{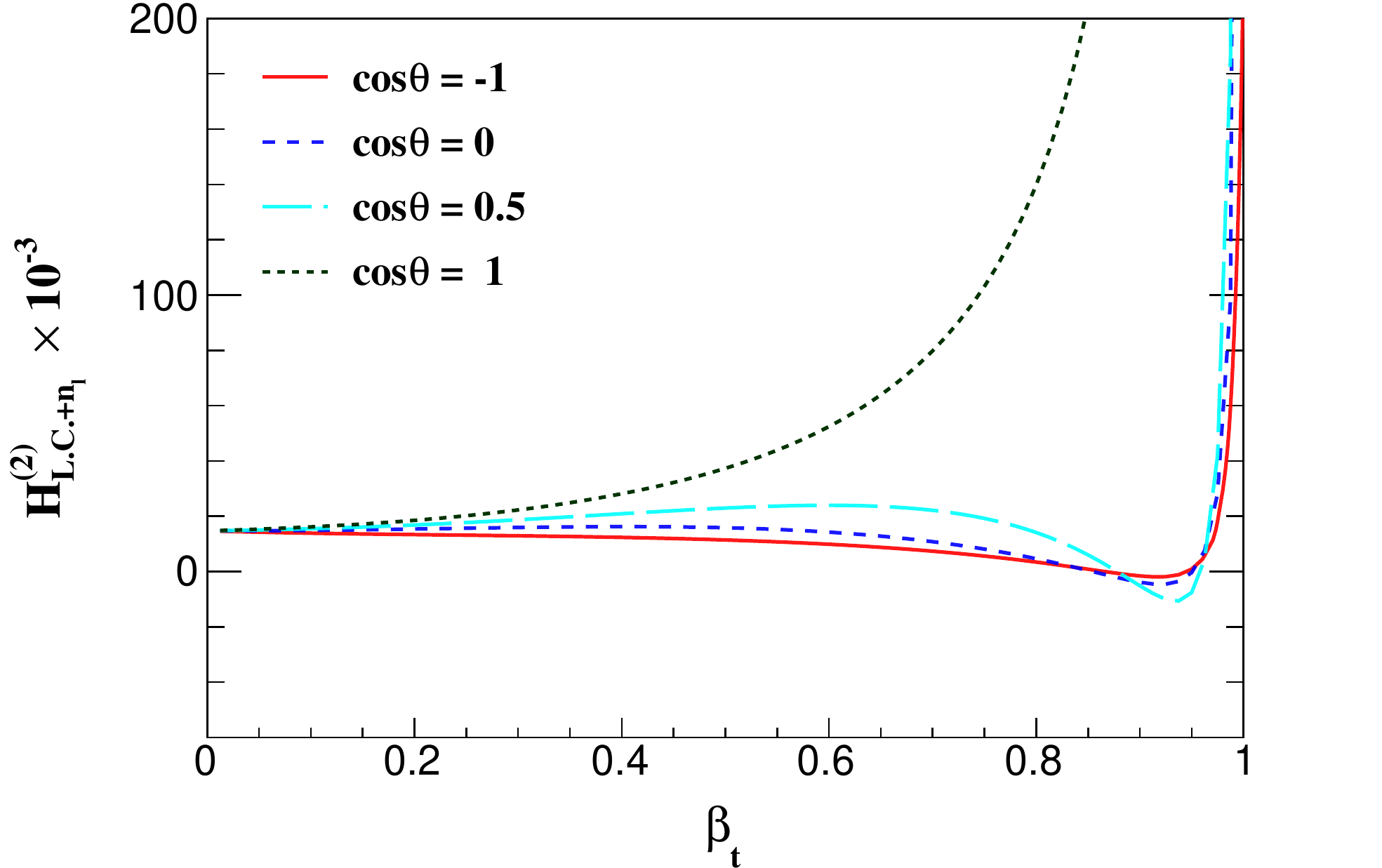}
    \includegraphics[width = 0.49 \textwidth]{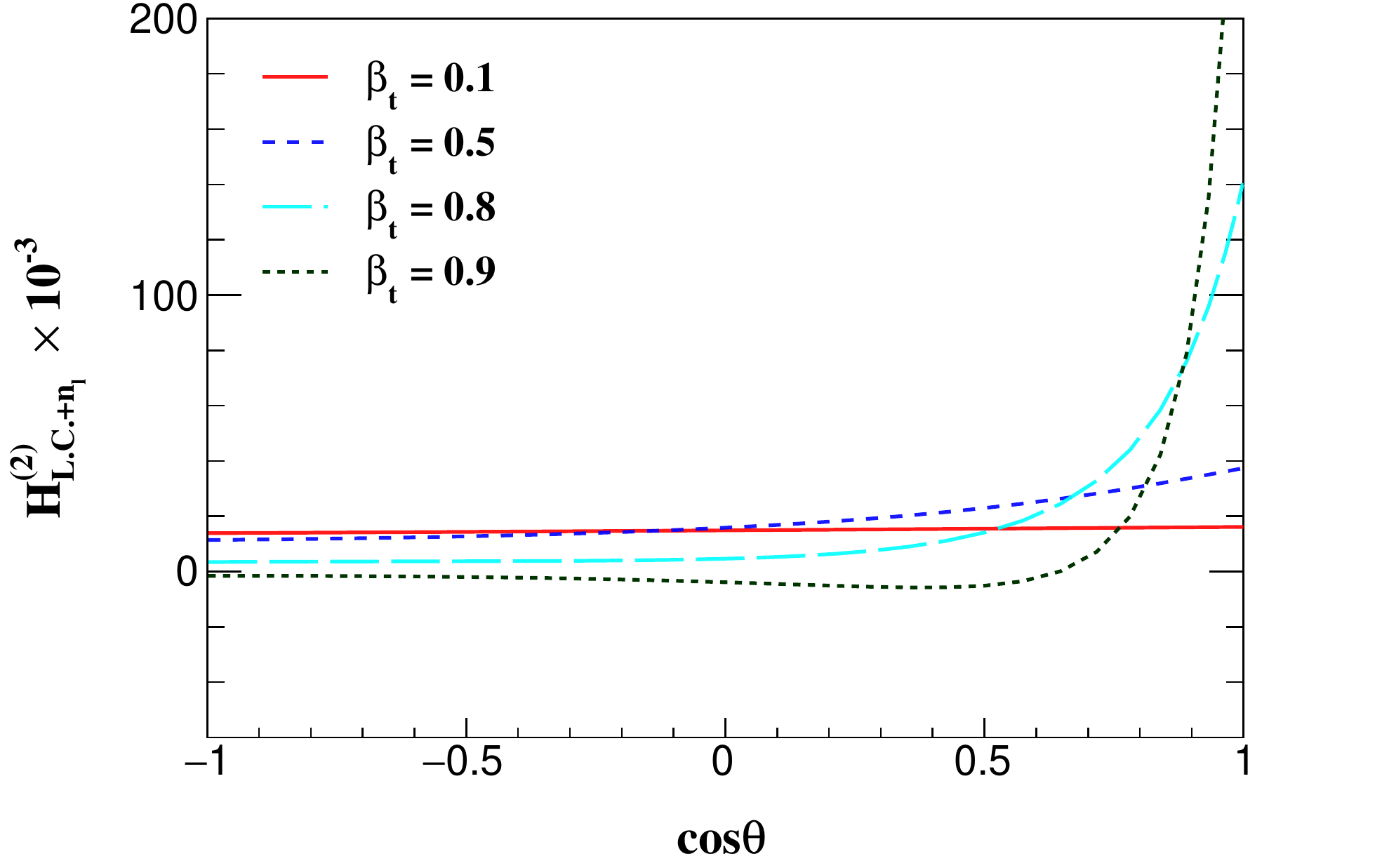}
    \caption{Leading color (top  panel) and the sum of leading color and light fermion-loop contribution  (bottom panel)   to the NNLO hard functions. }
    \label{fig:h2LC}
\end{figure}

\begin{figure}
    \centering
    \includegraphics[width = 0.49 \textwidth]{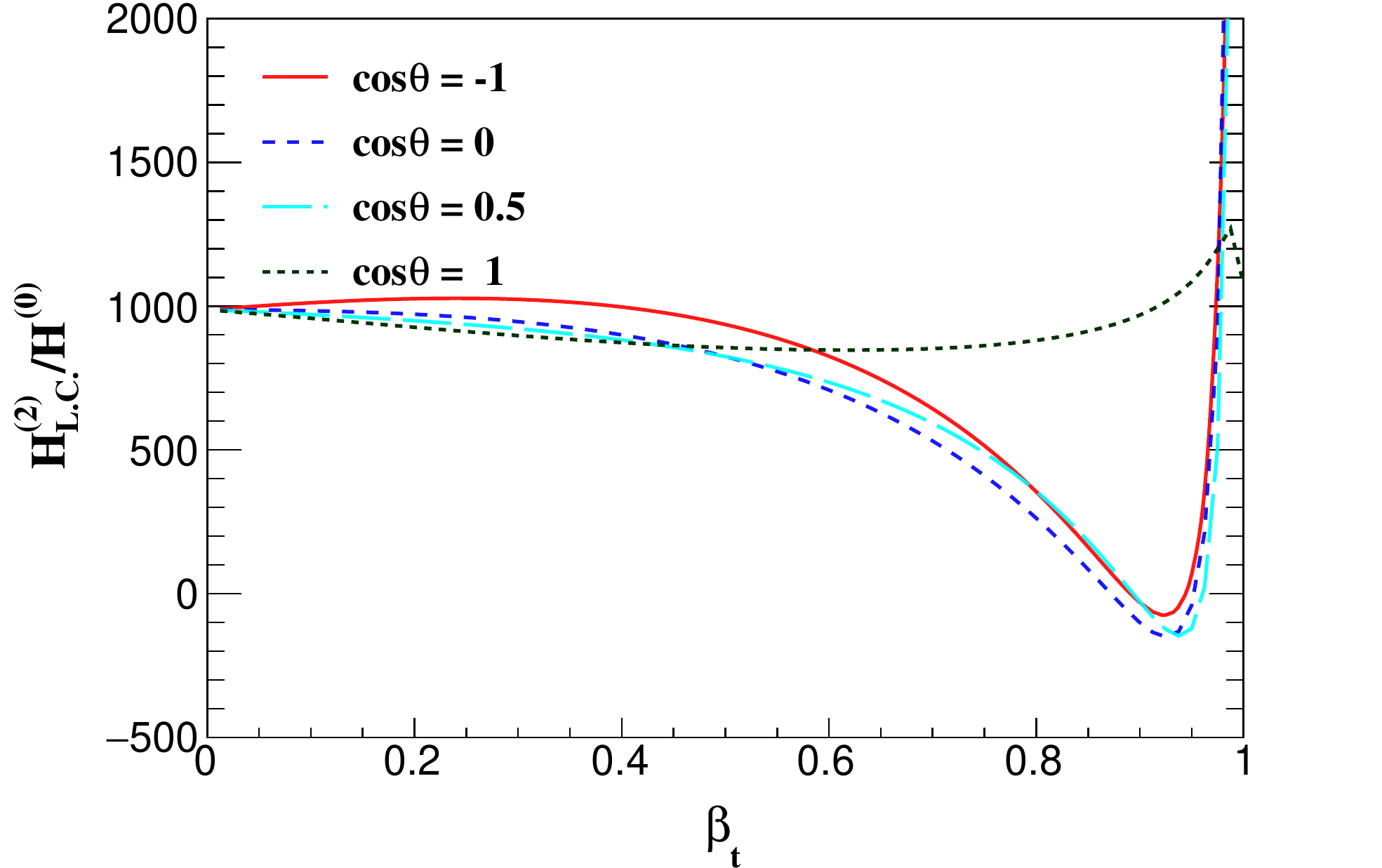}
    \includegraphics[width = 0.49 \textwidth]{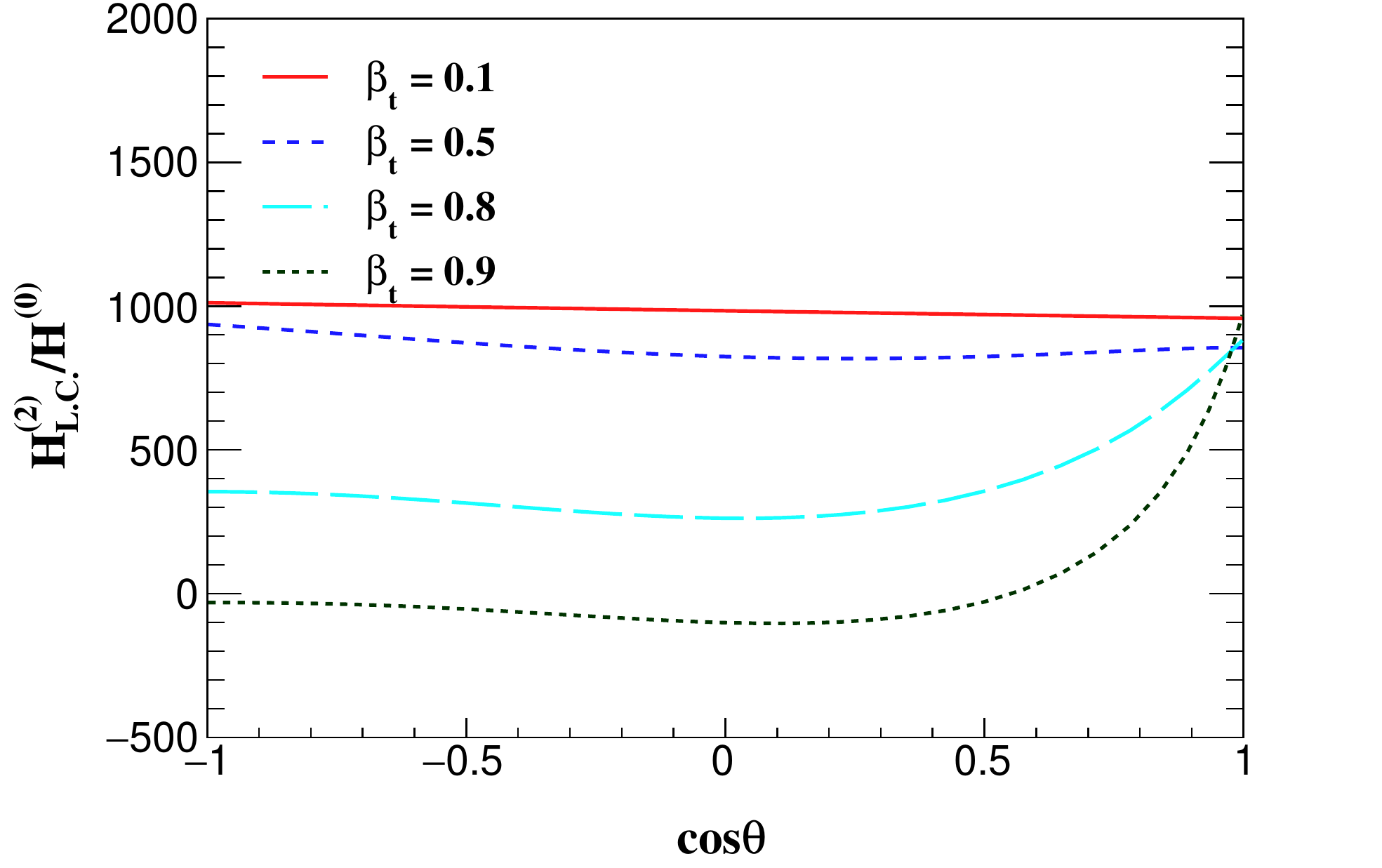}
    \includegraphics[width = 0.49 \textwidth]{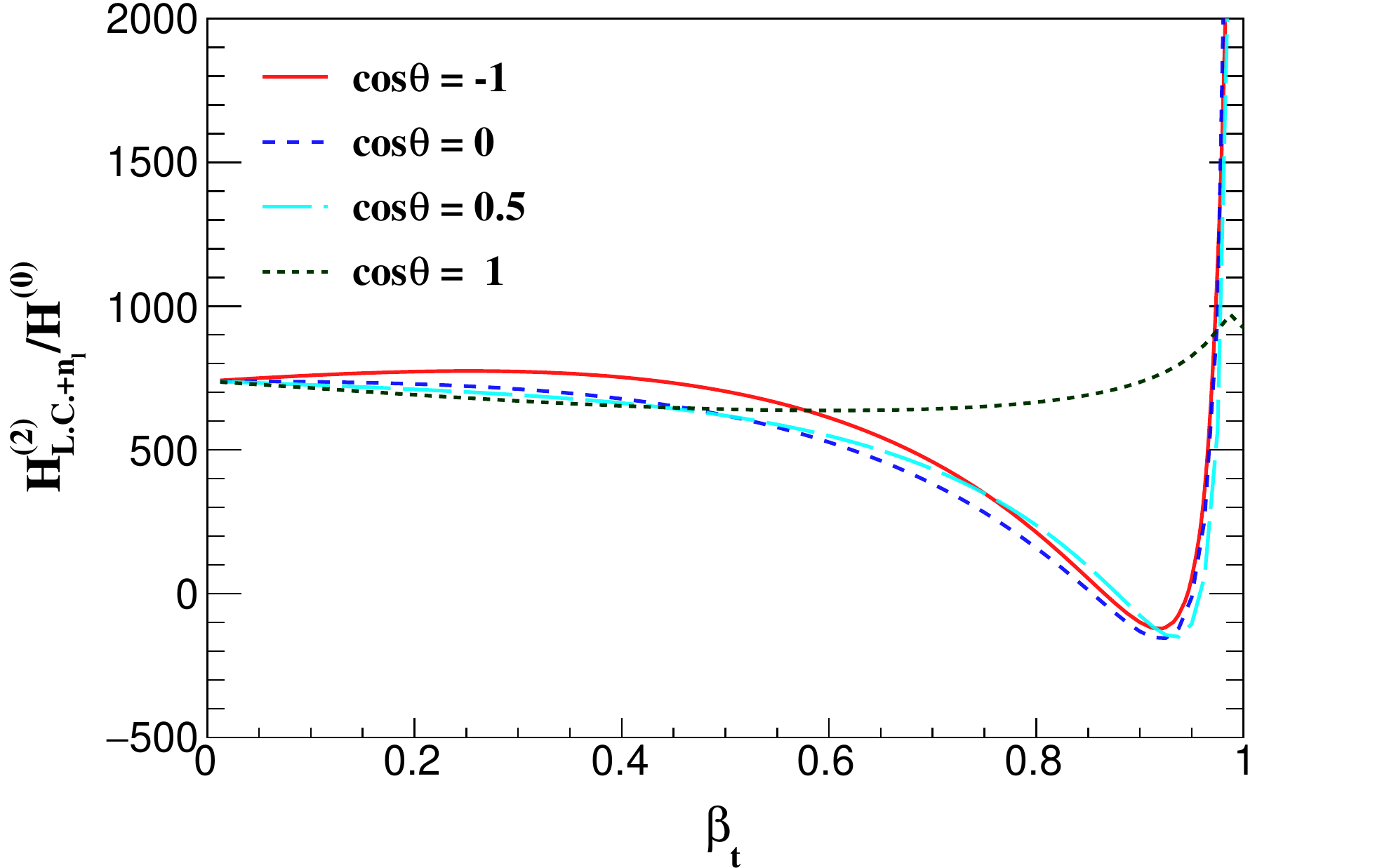}
    \includegraphics[width = 0.49 \textwidth]{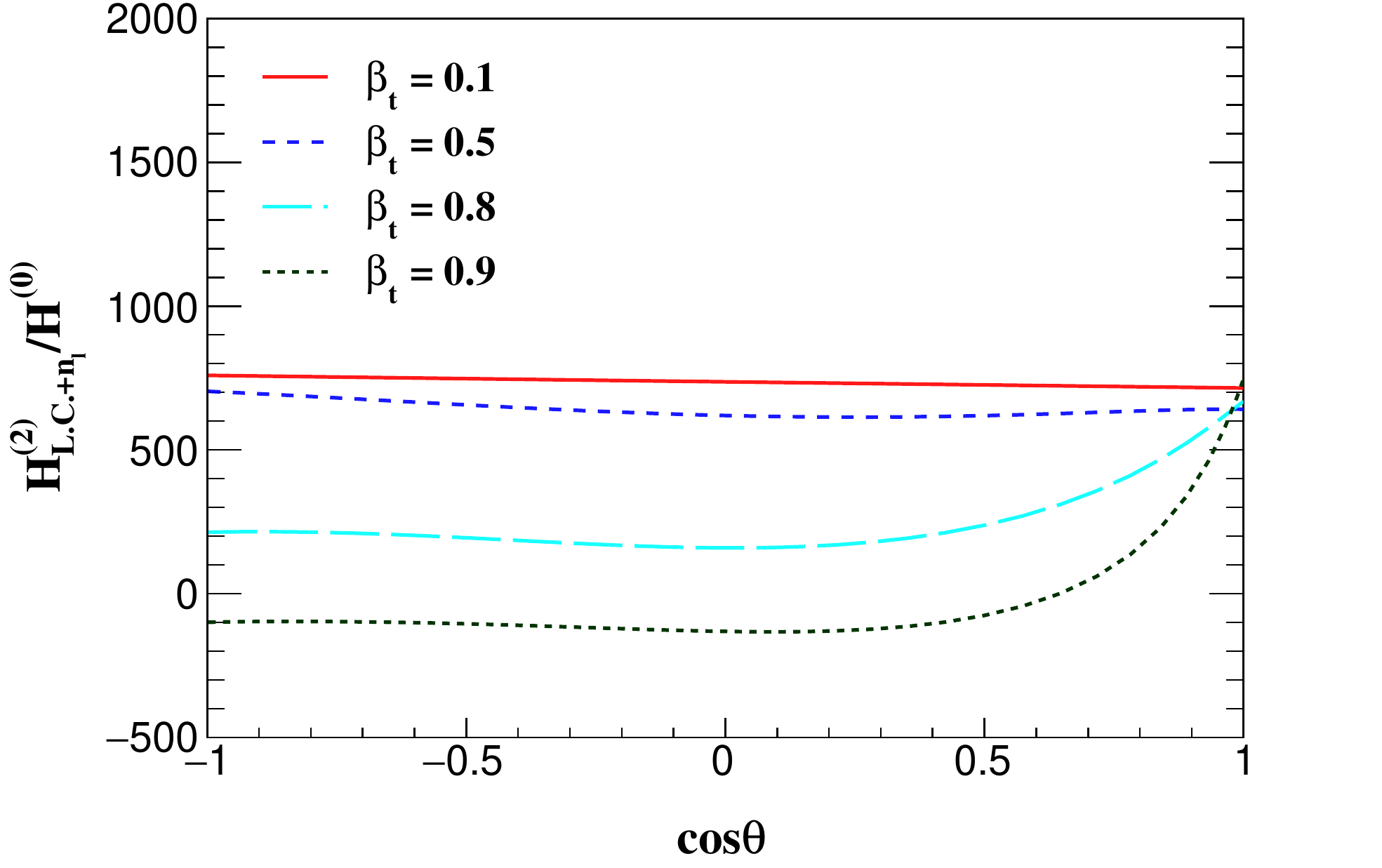}
    \caption{ Same as figure  \ref{fig:h2LC} but normalized by 
    $H^{(0)}$.}
    \label{fig:h2LCratio}
\end{figure}

Then we turn to the leading color NNLO  hard function. 
As shown in figure \ref{fig:h2LC},
it changes very slowly in the region with $\beta_t\le 0.7$ and $\cos\theta\le 0.6$.
For small $\beta_t$, this flat region extends to almost the whole $\cos\theta$ range.
But the leading hard function becomes divergent quickly  for large $\beta_t$ and general $\cos\theta$.
This reflects the fact that the amplitude develops new singularities in the limit $m_t\to 0$.
One can also observe the strongest divergence in the case of $\cos\theta=1$.
This behaviour is due to the tree-level propagator that contains $1/(1-\beta_t\cos\theta)$.
This kind of divergence should cancel in the ratio $H^{(2)}/H^{(0)}$, as displayed in figure \ref{fig:h2LCratio}.
However, the divergences arising from higher-order corrections, which are manifested in  the logarithmic terms $\ln(1-\beta_t^2)$, $\ln(1-\beta_t\cos\theta )$ and $\ln(1+\beta_t\cos\theta )$, survive in the ratio. 
Though they are numerically large, the hadronic cross section still drops dramatically as $\beta_t\to 1$ due to the overwhelming suppression of the parton distribution function \cite{Chen:2022ntw}.
Therefore, we do not need to worry about the divergent behavior in figure \ref{fig:h2LCratio}.

The light fermion loop may also be important due to the large value of $n_l$.
In figures \ref{fig:h2LC} and \ref{fig:h2LCratio},
we also show the numerical result of the hard function defined in eq.~(\ref{eq:h2lcnl}), which includes such additional contribution.
We find that the fermion-loop diagrams provide negative contributions, and decrease the leading color result by about $30\%$ in most of the phase space.
To be more specific and realistic, we perform the convolution with parton distribution functions \footnote{We choose CT14NLO \cite{Dulat:2015mca} in practical calculation, and the factorization and renormalization scales are set at $m_t$.} and integrate over all phase space, observing that the leading color and  $n_l$ dependent NNLO hard function  contribute  $5.4\%$ and $-1.4\%$, respectively, corrections to the LO cross section at the 13 TeV LHC.

\section{Conclusion} \label{sec:concl}

We have computed the two-loop virtual corrections to $tW$ production at hadron colliders, focusing on the leading color and light fermion-loop contributions.
Using the method of differential equations,
the results are obtained in terms of multiple polylogarithms.
After renormalization, we find that the IR divergences exhibit  a structure that is fully determined by the anomalous dimensions, which are known in the literature.
The finite part of the two-loop amplitude contributes to the hard function, which is an essential ingredient for a NNLO Monte Carlo calculation.
Numerical evaluation of the one-loop squared  amplitude confirms that the leading color result gives the dominant contribution.
The leading color NNLO  hard function shows very weak kinematic dependence in the region where the velocity $\beta_t$ of the top quark is small.
But it increases dramatically as $\beta_t$ approaches 1.
This behaviour is caused by two kinds of singularities.
The first kind of singularity disappears after normalization by the LO hard function, while the second still exists as logarithmic terms.
However, they do not have large impact on the hadronic cross section due to the suppression of the parton distribution function in the region of $\beta_t\to 1$.
The light fermion-loop contribution is negative, amounting to $30\%$ of the leading color result.
The sum of the leading color and light fermion-loop contribution to the NNLO hard function brings about $4\%$ correction to the LO cross section at the 13 TeV LHC.

Combined with the NNLO N-jettiness soft function \cite{Li:2016tvb,Li:2018tsq} and beam functions \cite{Gaunt:2014xga,Gaunt:2014cfa}, it is now promising to calculate the dominant NNLO QCD correction to the cross section of $tW$ production using the N-jettiness subtraction method \cite{Gao:2012ja,Boughezal:2015dva,Gaunt:2015pea}.

\section*{Acknowledgements}
This work was supported in part by the National Science Foundation of China under grant  Nos.12005117, 12075251, 12147154 and 12175048.  The work of L.D. and J.W. was also supported by the Taishan Scholar Foundation of Shandong province (tsqn201909011).

\bibliography{loop}
\bibliographystyle{JHEP}

\end{document}